%% file: main.tex
\documentclass[]{fairmeta}

\usepackage{hyperref}
\usepackage{url}

\usepackage[utf8]{inputenc} % allow utf-8 input
\usepackage[T1]{fontenc}    % use 8-bit T1 fonts
\usepackage{url}            % simple URL typesetting
\usepackage{booktabs}       % professional-quality tables
\usepackage{amsfonts}       % blackboard math symbols
\usepackage{nicefrac}       % compact symbols for 1/2, etc.
\usepackage{enumitem}
\usepackage{subcaption}
\usepackage{wasysym}

\usepackage{microtype}      % microtypography
\usepackage{graphicx}
\usepackage{adjustbox}
\usepackage{float}               % For float placement control

\usepackage{amsmath,amssymb,amsthm,bm,dsfont}
\usepackage{array, booktabs, makecell}
\usepackage{multirow}
\usepackage{cleveref}

\usepackage{pifont}

\usepackage[T1]{fontenc} % Ensures proper font encoding
\usepackage{courier}    % For Courier font
\usepackage{relsize}

\newcommand{\typing}[1]{\textsc{\smaller #1}}

\title{Brain-to-Text Decoding: \\
A Non-invasive Approach via Typing}

\author[1]{Jarod Lévy}
\author[2,3]{Mingfang (Lucy) Zhang}
\author[4,5]{Svetlana Pinet}
\author[1]{Jérémy Rapin}
\author[1]{Hubert Banville}
\author[1*]{Stéphane d'Ascoli}
\author[1*]{Jean-R\'emi King}

\affiliation[1]{Meta AI}
\affiliation[2]{École Normale Supérieure, Université PSL, CNRS}
\affiliation[3]{Hospital Foundation Adolphe de Rothschild}
\affiliation[4]{Basque Center on Cognition, Brain and Language, San Sebastian}
\affiliation[5]{Ikerbasque, Basque Foundation for Science, Bilbao}

\contribution[*]{equal contribution}

\abstract{Modern neuroprostheses can now restore communication in patients who have lost the ability to speak or move. However, these invasive devices entail risks inherent to neurosurgery. Here, we introduce a non-invasive method to decode the production of sentences from brain activity and demonstrate its efficacy in a cohort of 35 healthy volunteers. For this, we present Brain2Qwerty, a new deep learning architecture trained to decode sentences from either electro- (EEG) or magneto-encephalography (MEG), while participants typed briefly memorized sentences on a QWERTY keyboard. With MEG, Brain2Qwerty reaches, on average, a character-error-rate (CER) of 32\% and substantially outperforms EEG (CER: 67\%). For the best participants, the model achieves a CER of 19\%, and can perfectly decode a variety of sentences outside of the training set. While error analyses suggest that decoding depends on motor processes, the analysis of typographical errors suggests that it also involves higher-level cognitive factors. Overall, these results narrow the gap between invasive 
and non-invasive methods and thus open the path for developing safe brain-computer interfaces for non-communicating patients.

}

\date{\today}
\correspondence{\email{\{jarod,sdascoli,jeanremi\}@meta.com}}

\begin{document}

% \twocolumn
\maketitle
\input{figures/fig1.tex}
\section{Introduction}

% General motivation
The past decade has been marked by rapid progress in brain-computer interfaces (BCIs) for individuals who, after a brain lesion, have lost their ability to speak or communicate. 
% SOTA intracranial
In particular, several patients suffering from anarthria \citep{Metzger2022, moses2021neuroprosthesis}, Amyotrophic Lateral Sclerosis (ALS) \citep{willett2023neuroprosthesis}, or severe paralysis \citep{Hochberg2012} have now been able to produce full sentences via a neuroprosthesis, which records and decodes neural activity from motor regions of the brain. Originally limited to decoding small sets of linguistic features  \citep{herff2019generating, angrick2019speech, anumanchipalli2019speech, moses2021neuroprosthesis, card2024accurate}, words \citep{Metzger2022}, and gestures \citep{willett2021handwriting}, the recent development of AI models has improved the precision and rapidity of brain-to-text decoding to a point of enabling natural language production at rates close to normal speech \citep{metzger2023neuroprosthesis, wairagkar2024voice}.

% Challenge 1: invasive is dangerous and does not scale
However, such invasive neuroprostheses require a neurosurgical procedure, and thus expose patients to non-negligible risks of brain hemorrhage and infection \citep{chung2019highdensity, Bullard2020, Leuthardt2021, Baranauskas2014}. Additionally, maintaining functional cortical implants over extended time periods remains challenging~\citep{Fekete2023, Zhou2024, Yasar2024}. As a result, in their current form, \emph{invasive} BCIs are not easily scalable for diagnosing or restoring communication in the large groups of non- or poorly-responsive patients \citep{owen2006detecting, Claassen2019}.

% Challenge 2: non invasive is too noisy
\emph{Non-invasive} BCIs could potentially address this challenge. However, they are usually based on scalp electroencephalography (EEG), whose limited signal-to-noise ratio \citep{Mak2009} requires users to perform complex tasks. For example, EEG-based BCIs typically require individuals to maintain their attention on flickering stimuli \citep{abiri2019review} or to imagine moving their hand or foot over long time periods \citep{bodien2024cognitive} -- two tasks known to produce EEG patterns that can be relatively easily detected by a linear classifier. Even so, decoding performance remains moderate. For instance, a public BCI benchmark \citep{chevallier2024} using EEG achieves an accuracy of only 43.3\% on a four-class classification task with a motor imagery dataset \citep{Yi2014}. In sum, current non-invasive methods fall short of providing a fast and reliable BCI.

Two elements could address these challenges. First, magnetoencephalography (MEG), which measures the fluctations of magnetic fields elicited in the cortex, has higher signal-to-noise ratio than EEG \citep{Hamalaainen1993, Goldenholz2009, Baillet2017}. Second, deep learning models trained to reconstruct natural language from MEG signals in language \emph{comprehension} paradigms have recently demonstrated major improvements, especially in comparison to EEG \citep{defossez2023decoding}. Together, these elements thus indicate that with modern AI techniques, high-quality MEG signals and natural language tasks could be combined to decode the \emph{production} of language from non-invasive recordings of the brain.

% Approach
In this study, we introduce Brain2Qwerty, an AI model trained to decode text production from non-invasive recordings of brain activity (Fig. \ref{fig:approach}). For this, we tasked 35 participants to type briefly memorized sentences on a keyboard, while their brain activity was recorded with either EEG or MEG. We then train Brain2Qwerty, a three-stage deep neural network trained to decode text from these brain signals and evaluate it on both EEG (20 participants, 146K characters, 23K words and 4K sentences) and MEG recordings (20 participants, 193K characters, 30K words and 5K sentences).
Note that the present study does \emph{not} delve into \emph{how} the brain produce language during  typing. This neuroscientific issue is addressed in a companion paper \citep{lucy2025}.

% \begin{itemize}
%     \item Non-invasive techniques, such as electroencephalography (EEG) and magnetoencephalography (MEG), require users to engage in prolonged and demanding tasks like sustained attention or motor imagery \cite{bodien2024cognitive}.
%     \item Example: The P3-Speller allows participants to spell letters by focusing on flickering stimuli on a screen but is slow and demanding \cite{abiri2019review}.
%     % \item Extensive literature exists on EEG-based brain decoding; however, most models heavily rely on teacher-forcing during evaluation. This reliance limits their applicability in real-world scenarios where such controlled conditions are unavailable \cite{wang2022open, duan2023dewave}.
% \end{itemize}

%%%

% \begin{table*}
%     \footnotesize
%     \centering
%     \begin{tabular}{lccccccccc}  % Corrected to 10 columns (one less 'c')
%     \hline
%     Dataset & Subjects & \makecell{Blocks} & Time (h) & \makecell{Avg\\Duration (h)} & Words & \makecell{Unique\\Words} & \makecell{Unique\\Sentences} & Keystrokes & Sensors \\ \hline
%     EEG & 20 & 62 & 17.7 & 0.88 & 23,040 & 351 & 128 & 146,551 & 64  \\
%     MEG & 19 & 81 & 21.5 & 0.93 & 30,626 & 351 & 128 & 193,094 & 306  \\ \hline
%     \end{tabular}
%     \caption{\textbf{Dataset Characteristics.}
%     %This table summarizes the key attributes of the datasets analyzed in this study.
%     }
%     \label{tab:dataset}
% \end{table*}

\section{Results}

% \begin{figure*}[!ht]
%     \centering 
%     \hspace{-0.8cm}
%     \includegraphics[width=0.7\textwidth]{figures/fig2.pdf}
%     \caption{
%         \textbf{Linear Decoding Performance.} \\
% A linear ridge classifier model is trained to predict either the hand used to type (A) or the character typed (B). Results for EEG are shown in blue, while MEG results are displayed in green. The thin colored lines represent chance level. Intervals where decoding accuracy is significantly above chance (p < 0.05) are marked with a star.
%     }
%     \label{fig:linear_perf}
% \end{figure*}

% \begin{figure}[!ht]
%     \hspace{-0.8cm}
%     \includegraphics[width=0.5\textwidth]{figures/fig3.2.pdf}
%     \caption{
%         \textbf{Model Decoding Performance.} \\
% Comparison of decoding performance across \textbf{EEG (left)} and \textbf{MEG (right)} for baseline models (Linear and EEGNet \cite{lawhern2018EEGNet}) and our proposed model (Ours). Each point represents a participant and the mean value is displayed. Statistical significance (one-sided Wilcoxon tests across the participants) are denoted as follows: p < 0.05 (*), p < 0.01 (**), and p < 0.001 (***).
%     }
%     \label{fig:main_results}
% \end{figure}

\input{figures/fig2.tex}

\subsection{Linear Decoding}
To verify that our typing protocol leads to the expected brain responses, we first focus on the differences in evoked responses elicited by left- and right-handed key presses (Fig. \ref{fig:model_perf} A-B). The resulting topographies are typical of those associated with motor activity in the cortex \citep{Donner2009}. In addition, we trained a linear ridge classifier per subject to categorize left- vs right-handed responses at each time-sample relative to key presses. The classification accuracy peaks t=40\,ms after the key press (Fig. \ref{fig:model_perf}). MEG achieves a peak accuracy of 74$\pm$1.3\% ($\pm$ reports standard-error-of-the-mean (SEM) across subjects), significantly outperforming EEG which achieves 64$\pm$0.8\% (Mann-Withney U test: p<$10^{-7}$). We then trained the same model to classify the keys pressed. Character accuracy peaks around the same time, reaching a value of 22$\pm$0.8\% for MEG and 16$\pm$0.5\% for EEG, significantly above the chance level (14\%). Overall, these findings confirm that the present protocol leads to the expected brain responses to key presses 
\citep{Pinet2020}.

\subsection{Brain2Qwerty Performance}
We next trained Brain2Qwerty, a new deep learning architecture, to decode individual characters from these M/EEG signals (see Methods in \cref{subsection:decoder}) and evaluated both the hands-error-rate (HER) and the character-error-rate (CER).
Brain2Qwerty achieves a CER of 32$\pm$0.6\% with MEG and 67$\pm$1.5\% with EEG. 
This performance reflects a substantial difference across recording devices (p<$10^{-8}$). The best and worst EEG subjects reach a CER of 61$\pm$2.0\% across sentences and 71$\pm$2.3\%, respectively. Similarly, the best and worst MEG subjects reach a CER of 19$\pm$1.1\% and 45$\pm$1.2\%, respectively. 

% For the best-performing subject, the error rate ranges from 0.0 to 0.6, while for the worst-performing subject, it ranges from 0.2 to 0.8. Despite this variability, these results demonstrate that even for the subject with the highest error rates, the model is still capable of decoding sentences with reasonable accuracy in most cases. 

\subsection{Comparing Brain2Qwerty to Baseline Models}
How does Brain2Qwerty perform in comparison to classic baseline architectures? To address this issue we trained a linear model as well as EEGNet --  a popular architecture used in BCIs \citep{lawhern2018EEGNet} -- with the same approach, and compared their decoding performance to Brain2Qwerty's with a Wilcoxon test across subjects (Fig. \ref{fig:model_perf}~E-H).
EEGNet outperforms the linear model on both HER (p=0.008) and CER (p<$10^{-4}$) for MEG  -- although only on HER for EEG (p=0.03). %, hence confirming that deep learning is effective at extracting non-linear patterns embedded in the M/EEG signals
However, EEGNet remains less effective than our model, which achieves, in comparison, a 1.14-fold improvement in CER with EEG (p<$10^{-5}$) and a 2.25-fold improvement for MEG (p<$10^{-6}$), respectively.

\subsection{Brain2Qwerty Ablations}
To validate our design choices, we then retrained different ablated versions of our model. Specifically, we re-trained and evaluated (i) the Convolutional Module (i.e. no transformer, no language model) and (ii) the Conv+Transformer (i.e. without Language Model) with the same hyperparameters. 
%
% The difference in CER between the baseline models and our pipeline is illustrated by a p of $10^{-6}$ for EEG and $10^{-7}$ for MEG (Wilcoxon signed-rank text).
%
%In \cref{fig:model_perf}, we initially present the performance of our model on a conventional Brain-Computer Interface (BCI) task, specifically hand recognition. %Notably, our model achieves character-error-rates (CERs) of 26\% and 9\% for this task.
%
The Convolutional Module alone outperforms EEGNet both on EEG (HER: p=0.009, CER: p=0.03) and MEG (HER: p<$10^{-5}$, CER: p<$10^{-6}$). %This result suggests that the subject-embedding layer of our Convolutional Module is important to maximize decoding performance. 
Adding the transformer only appears beneficial to CER, both for EEG (p<$10^{-4}$) and MEG (p<$10^{-6}$). 
Finally, the use of a Language Model module leads to an additional improvement of the CER of EEG (p<$10^{-5}$) and MEG (p<$10^{-6}$).
Overall, these results show that the sentence-level contextualization provided by the transformer together with the leverage of natural language's statistical regularities effectively improves the decoding of individual characters. 

\input{figures/fig3.tex}

\subsection{Analyses of Decoded Sentences}
The CER of all sentences for three representative participants recorded with MEG along with two example sentences from these subjects are displayed in Fig. \ref{fig:performance_sentence}. 
% (CERs of 0.28, 0.31 and 0.32). %In this example, we present the decoded sentences for a single trial per subject. 
More decoding examples show that several sentences can be perfectly decoded for MEG (Tab. \ref{tab:combined_decoded_sentences}, right). Interestingly, some of these examples show that Brain2Qwerty's language model can correct the typographical errors of the participant. For example, \textsc{el beneficio supera los riesgos} was perfectly decoded, even though the participants typed: \textsc{ek benefucui syoera kis ruesgis}. In comparison, the poor EEG decoding (Tab. \ref{tab:combined_decoded_sentences}, left) rarely leads to comprehensible text. % respectively CERs of 0.32, 0.42 and 0.43 underscoring again the superior quality of results obtained with MEG.
%
%Shorter words, such as determiners, are consistently well-predicted. However, decoding longer words proves more challenging, especially when the output of the Transformer module significantly deviates from the intended word. In such cases, the N-gram model tends to predict a word that is far from the correct one. This highlights a trade-off inherent to the N-gram model: while it performs well when the Transformer provides reasonably accurate predictions, its performance deteriorates when the input is highly erroneous.
%
%We presents two cherry-picked 
% Other representative examples of sentences decoded with each ablation of our model are displayed in \cref{tab:step_by_step_decoded_sentences}. 
%In the first example, the CER starts at 0.52 after the CNN module, decreases to 0.16 after the Transformer, and ultimately reaches 0.0 after the final N-gram module. Similarly, in the second example, the CER begins at 0.34 after the CNN, reduces to 0.16 following the Transformer, and is perfectly decoded by the end of the pipeline.
Consistent with the statistical effects reported earlier, the examples in Tab. \ref{tab:step_by_step_decoded_sentences} highlight the impact of each module of our model, which together leads to perfect decoding after the language model. %This last step is particularly striking, as it can reconstruct complex words like \textit{teorías} from highly erroneous intermediate predictions such as \textit{geoolas}. Overall, these results showcase the feasibility of leveraging context and prior linguistic knowledge to improve brain-to-text decoding.

\begin{table*}
    \centering
    \begin{minipage}{0.50\textwidth}
        \centering
        \textbf{EEG}\\\vspace{.2cm}
        \begin{tabular}{p{0.12\textwidth} p{0.88\textwidth}}
    \textbf{Read:} & \typing{la ciencia de la idea rompe la vision} \\
    \textbf{Typed:} & \typing{la ciencia de la idea rompe la \underline{b}ision} \\
    \textbf{Decode:} & \typing{\textcolor[rgb]{0.204, 0.596, 0.859}{la ciencia de la idea} \textcolor[rgb]{0.882, 0.071, 0.188}{las mas de es}\textcolor[rgb]{0.204, 0.596, 0.859}{o}\textcolor[rgb]{0.882, 0.071, 0.188}{s}}\\
    
    \addlinespace
    
    \textbf{Read:} & \typing{el procesador ejecuta la instruccion} \\
    \textbf{Typed:} & \typing{\underline{orden}ador ejecuta la instruccion} \\
    \textbf{Decode:} & \typing{\textcolor[rgb]{0.882, 0.071, 0.188}{las c}\textcolor[rgb]{0.204, 0.596, 0.859}{o}\textcolor[rgb]{0.882, 0.071, 0.188}{rrida peri}\textcolor[rgb]{0.204, 0.596, 0.859}{ta la instruccion}} \\
    
    \addlinespace
    
    \textbf{Read:} & \typing{la presencia de los tipos impone los retos} \\
    \textbf{Type:} & \typing{la presencia de los tipos impone los retos} \\
    \textbf{Decode:} & \typing{\textcolor[rgb]{0.204, 0.596, 0.859}{la }\textcolor[rgb]{0.882, 0.071, 0.188}{declarad}\textcolor[rgb]{0.204, 0.596, 0.859}{a de los }\textcolor[rgb]{0.882, 0.071, 0.188}{cel}\textcolor[rgb]{0.204, 0.596, 0.859}{os }\textcolor[rgb]{0.882, 0.071, 0.188}{eran a }\textcolor[rgb]{0.204, 0.596, 0.859}{los }\textcolor[rgb]{0.882, 0.071, 0.188}{ac}\textcolor[rgb]{0.204, 0.596, 0.859}{tos}} \\
    
    \addlinespace
    
    \end{tabular}
    \end{minipage}
    \hfill
    \begin{minipage}{0.46\textwidth}
        \centering
        \textbf{MEG}\\\vspace{.2cm}
        \begin{tabular}{p{0.12\textwidth} p{0.88\textwidth}}            
            \textbf{Read:} & \typing{la silla ocasiona las lesiones }\\
            \textbf{Type:} & \typing{la silla ocasio\underline{m}a las lesio\underline{m}es }\\
            \textbf{Decode:} & \typing{\textcolor[rgb]{0.204, 0.596, 0.859}{la silla ocasiona las lesiones} }\\
            
            \addlinespace
            
            \textbf{Read:} & \typing{las teorias reducen los numeros }\\
            \textbf{Type:} & \typing{las teorias reducen los numeros }\\
            \textbf{Decode:} & \typing{\textcolor[rgb]{0.204, 0.596, 0.859}{las teorias reducen los numeros} }\\
            
            \addlinespace

            \textbf{Read:} & \typing{el beneficio supera los riesgos }\\
            \textbf{Type:} & \typing{e\underline{k} benef\underline{u}c\underline{ui} s\underline{yo}era \underline{ki}s r\underline{u}esg\underline{i}s }\\
            \textbf{Decode:} & \typing{\textcolor[rgb]{0.204, 0.596, 0.859}{el beneficio supera los riesgos} }\\
            \addlinespace
        \end{tabular}
    \end{minipage}
    \caption{\textbf{Examples of best-decoded sentences across subjects for EEG (left) and MEG (right) data.} \\
    Correct characters are highlighted in \textcolor[rgb]{0.204, 0.596, 0.859}{blue}, mistakes in \textcolor[rgb]{0.882, 0.071, 0.188}{red}, and typing errors \underline{underlined}. Note that correct vs incorrect spaces are not visualized here.\\
    % \textbf{(Right)} Step-by-step decoding using CNN, CNN+Transformer, and CNN+Transformer+Ngram.
    }
    \label{tab:combined_decoded_sentences}
\end{table*}

\begin{table*}
    \centering
    \begin{minipage}{0.45\textwidth}
        \centering
        \begin{tabular}{p{0.05\textwidth} p{0.85\textwidth}}
            \textbf{Read:} & \hspace{1.5cm} \typing{las teorias reducen los numeros} \\
            \textbf{Typed:} & \hspace{1.5cm} \typing{las teorias reducen los numeros} \\
            \textbf{Conv:} & \hspace{1.5cm} \typing{\textcolor[rgb]{0.204, 0.596, 0.859}{las t}\textcolor[rgb]{0.882, 0.071, 0.188}{angp}\textcolor[rgb]{0.204, 0.596, 0.859}{as re u}\textcolor[rgb]{0.882, 0.071, 0.188}{dn}\textcolor[rgb]{0.204, 0.596, 0.859}{n}\textcolor[rgb]{0.882, 0.071, 0.188}{d}\textcolor[rgb]{0.204, 0.596, 0.859}{l s}\textcolor[rgb]{0.882, 0.071, 0.188}{lindiis}} \\
            \textbf{Conv+Trans:} & \hspace{1.5cm} \typing{\textcolor[rgb]{0.204, 0.596, 0.859}{las} \textcolor[rgb]{0.882, 0.071, 0.188}{g}\textcolor[rgb]{0.204, 0.596, 0.859}{eo}\textcolor[rgb]{0.882, 0.071, 0.188}{ol}\textcolor[rgb]{0.204, 0.596, 0.859}{as red}\textcolor[rgb]{0.882, 0.071, 0.188}{i}\textcolor[rgb]{0.204, 0.596, 0.859}{cen los nume}\textcolor[rgb]{0.882, 0.071, 0.188}{i}\textcolor[rgb]{0.204, 0.596, 0.859}{os}}\\
            \textbf{Brain2Qwerty:} & \typing{\hspace{1.5cm} \textcolor[rgb]{0.204, 0.596, 0.859}{las teorias reducen los numeros}}\\
    \end{tabular}
    \end{minipage}
    \hfill
    \begin{minipage}{0.45\textwidth}
        \centering
        \begin{tabular}{p{0.05\textwidth} p{0.85\textwidth}}
            \textbf{Read:} & \hspace{1.5cm} \typing{el beneficio supera los riesgos} \\
            \textbf{Typed:} & \hspace{1.5cm} \typing{el be\underline{m}eficio supera los ries\underline{f}os} \\
            \textbf{Conv:} & \hspace{1.5cm} \typing{\textcolor[rgb]{0.204, 0.596, 0.859}{el} \textcolor[rgb]{0.882, 0.071, 0.188}{g}\textcolor[rgb]{0.204, 0.596, 0.859}{e}\textcolor[rgb]{0.882, 0.071, 0.188}{f}\textcolor[rgb]{0.204, 0.596, 0.859}{e}\textcolor[rgb]{0.882, 0.071, 0.188}{d}\textcolor[rgb]{0.204, 0.596, 0.859}{i}\textcolor[rgb]{0.882, 0.071, 0.188}{s}\textcolor[rgb]{0.204, 0.596, 0.859}{io su}\textcolor[rgb]{0.882, 0.071, 0.188}{i}\textcolor[rgb]{0.204, 0.596, 0.859}{era} \textcolor[rgb]{0.882, 0.071, 0.188}{n}\textcolor[rgb]{0.204, 0.596, 0.859}{o}\textcolor[rgb]{0.882, 0.071, 0.188}{a} \textcolor[rgb]{0.204, 0.596, 0.859}{ries}\textcolor[rgb]{0.882, 0.071, 0.188}{tii} }\\
            \textbf{Conv+Trans:} & \hspace{1.5cm} \typing{\textcolor[rgb]{0.204, 0.596, 0.859}{el} \textcolor[rgb]{0.882, 0.071, 0.188}{g}\textcolor[rgb]{0.204, 0.596, 0.859}{enefic}\textcolor[rgb]{0.882, 0.071, 0.188}{on} \textcolor[rgb]{0.882, 0.071, 0.188}{c}\textcolor[rgb]{0.204, 0.596, 0.859}{upera los riesgo}\textcolor[rgb]{0.882, 0.071, 0.188}{o} }\\
            \textbf{Brain2Qwerty:} & \hspace{1.5cm} \typing{\textcolor[rgb]{0.204, 0.596, 0.859}{el beneficio supera los riesgos} }\\
            \addlinespace
        \end{tabular}
    \end{minipage}
    \caption{\textbf{Example of decoded sentences across ablations for MEG data.} 
    Color coding identical to Tab. \ref{tab:combined_decoded_sentences}.
    %Correct characters are highlighted in \textcolor[rgb]{0.204, 0.596, 0.859}{blue}, mistakes in \textcolor[rgb]{0.882, 0.071, 0.188}{red}, and typing errors are \underline{underlined}. 
    \\
    % \textbf{(Right)} Step-by-step decoding using CNN, CNN+Transformer, and CNN+Transformer+Ngram.
    }
    \label{tab:step_by_step_decoded_sentences}
\end{table*}

\subsection{Impact of Word Type and Frequency}

%In this section, we analyze the factors impacting the performance of our model. %Each analysis is conducted on the test set of each participant and averaged across participants. Results are reported in \cref{fig:model_analysis}.

To test whether Brain2Qwerty decodes words irrespectively of their grammatical type, we evaluated the CER for each part-of-speech (POS) categories separately (Fig. \ref{fig:model_analysis}A). All POS categories are significantly better decoded than chance, with determiners exhibiting a remarkably low CER (17$\pm$1.9\%). This phenomenon may be due to two factors: their short length and their high frequency. %However, WER is relatively stable across all categories, as long words benefit more from the language model correction step.
To formally test this hypothesis, we first analyzed the impact of word frequency on CER (Fig. \ref{fig:model_analysis}B). The results confirm that frequent words are better decoded than rare words (p=$10^{-7}$). Interestingly, we verify that the words absent from the training set (out-of-vocabulary, OOV) can also be decoded, although with a relatively poor CER (68$\pm$2.1\%). Note that this may be due to the fact that on a random partition of train/validation/test splits, OOV words tend to be rare words.

Second, we evaluated whether the frequency of each character also impacts decoding. The results show a significant correlation between character frequency and decoding accuracy: R=0.85, p < $10^{-8}$ (Fig. \ref{fig:model_analysis}C). Rare characters, such as "z," "k," and "w" in Spanish, are not decoded above chance level but only account for 0.08\%, 0.08\%, and 0.05\% of the characters in our sentences. These results suggest that the number of repetitions (of words and characters) encountered during training directly affects performance. 

To confirm this, we explore how decoding performance scales with the amount of data. For this, we re-trained our model on uniformly sampled subsets of the training set (Fig. \ref{fig:model_analysis}D). Our results show that CER decreases as a function of the amount of training data: R=0.93, p<$10^{-7}$.

\input{figures/fig4.tex}

\input{figures/fig5.tex}

\subsection{Impact of Keyboard Layout}
% hypothesis
If Brain2Qwerty relies on brain activity from the motor cortex (as opposed to some amodal representations of language), then we expect its decoding errors to relate to the specific layout of the QWERTY keyboard.
% supervised
To test this hypothesis, we evaluated the confusion patterns of incorrectly predicted characters by analyzing the keyboard distance between decoded and actual key presses. %Specifically, for each key, we identify all instances of misclassifications, categorize them based on their corresponding keyboard distances (defined as the minimum number of steps required to transition from one key to another). We then normalize across the keyboard distances. 
The results show a strong Pearson correlation between physical distance and confusion rate: R=0.73, p=0.02 (Fig. \ref{fig:motor_analysis}A). 
%This finding is intuitive, as participants are more likely to make errors with keys located near one another on the keyboard. 
% clustering
To complement this analysis, we further perform a clustering of the last-layer embeddings of the convolutional module using scikit-learn's K-means clustering algorithm. %  straightforward algorithm like K-means. 
When trained on two clusters, this unsupervised model fully separates the left-hand and right-hand keys. When using up to 10 clusters, the resulting partition remains consistent with the keyboard layout (Fig. \ref{fig:motor_analysis}B). This shows that the spatial layout of the keyboard is well encoded in the high-dimensional representations learned by our model.
% conclusion
Overall, these results show that the decoder's mistakes tend to be confused with the keys that are physically close to the target letter on the QWERTY keyboard, suggesting that the decoder primarily relies on motor representations.

\subsection{Impact of Typing Errors}
Our protocol does not allow participants to correct their mistakes. Typing errors account for 3.9\% of the keystrokes and are present in 65\% of the sentences. Furthermore, they are associated with a specific behavior (Fig. \ref{fig:motor_analysis}C): the time taken to type a character -- as measured by the inter-key interval -- doubles between correctly- (50$\pm$7\,ms) and incorrectly-typed characters (114$\pm$12\,ms, p=$10^{-7}$). This well-known phenomenon \citep{logan2010cognitive} likely reflects hesitation or monitoring of mistakes.  
%For sentences containing at least one typing mistake, the average number of typos per sentence is 2.8. 
To evaluate the impact of typing error on decoding performance, we separately evaluated CER for correctly- and incorrectly-typed characters (Fig. \ref{fig:motor_analysis}D). %Typing mistakes can be categorized as either additions or substitutions, both types of errors are considered. %To align the true sentence with the typed sentence, we use the \textit{SequenceMatcher} from the \texttt{difflib} Python library.
%
%Typing mistakes are significantly more frequent at the ends of sentences (p = $10^{-6}$) as shown in panel A. 
%
% 
The results show that with our Conv+Trans model, correctly-typed characters lead to a better CER (38\%) than incorrectly-typed characters (65\%, p=$10^{-7}$). This result, however, may be partly driven by the contextualization enabled by the transformer. 
To minimize the impact of sentence context on this error analysis, we thus evaluate the performance of the Convolutional Module (Fig. \ref{fig:motor_analysis}D, right). Again, correctly-typed characters leads to a better CER (52\%) than incorrectly-typed characters (71\%, p=$10^{-7}$). This result suggests that decoding performance diminishes when motor processes are inaccurately executed.

\section{Discussion}
% Summary / Contribution
The present study introduces a new method to decode the production of sentences from non-invasive brain recordings. With MEG, our Brain2Qwerty model achieves a character-error-rate (CER) of 32$\pm$0.6\% on average across subjects, with the best-performing participants reaching a CER as low as 19\%.
Our analyses indicate that this decoding benefits from two main factors. First, the use of MEG signals instead of EEG signals resulted in a two-fold improvement. Second, our deep learning architecture, combined with a pretrained character-level language model, substantially outperforms standard models.

%Projecting the embeddings from the last layer onto a QWERTY keyboard, as shown in \ref{fig:typing_analysis}, reveals clear patterns, suggesting that a significant portion of the model's representations can be attributed to hand movements. 
%It is not clear whether the analysis of typographical mistakes in \ref{fig:typo_analysis} could be explained by hesitation in hand movements during errors or may indicate more complex encoding patterns, such as the participant's underlying intentions.

% Comparison to current non-invasive decoding
This work directly stems from the recent progress in decoding natural language from non-invasive recordings of brain activity. In particular, \citet{defossez2023decoding} showed that the perception of natural speech segments could be decoded, from MEG signals, with up to 41\% top-10 accuracy (chance level=0.1\%). Similarly, \citet{tang2023semantic} showed that the meaning of perceived sentences could be decoded from functional Magnetic Resonance Imaging (fMRI). Our Brain2Qwerty model shares several elements with these approaches, notably through the use of both a subject layer \citep{defossez2023decoding} and the use of a language model \citep{tang2023semantic}, although here restricted to a pretrained 9-gram character-level model. However, these two studies, which focus on decoding the \emph{perception} of language rather than its \emph{production}, remained limited in their downstream clinical applications.

% Recently, \cite{sivakumar2024emgqwerty} introduced a method for decoding EMG signals during a typing task. The signals were recorded directly from the wrist, comprising 1,135 sessions, 108 participants, and a total of 346 hours of data. EMG signals are less noisy than MEG signals due to their direct recording from the wrist, which reduces artifacts, and their distinctive waveform characterized by clear spikes. By fine-tuning the model on individual participants, the method achieved an impressive error rate of just 7\%. While EMG signals appear highly suitable for decoding, they are inapplicable to individuals with severe motor impairments. 

Studies that directly decode text production from non-invasive recordings remain rare. For example, \citet{crell2024decoding} employed EEG to decode only 10 letters and achieved a character-error-rate of 75.8\%, substantially higher than our 68\% in the EEG setting with 29 characters. Similarly, EEG-based BCI benchmarks currently emphasize the constraints of poor signal quality and the variability across subjects \citep{chevallier2024}. Our EEG findings are consistent with these observations. Beyond its mere metric performance, our approach is more efficient than traditional protocols used in non-invasive BCIs, such as the P300-speller \citep{marchetti2014effectiveness}, SSVEP \citep{cheng2002design}, and fMRI-localizer \citep{owen2006detecting}. These methods have historically relied on handcrafted signal processing and shallow classifiers. In contrast,  our approach is based on a task that is comparatively easier to use.

% Comparison to invasive methods
While the decoding performance of Brain2Qwerty narrows the gap between non-invasive and invasive BCIs, this gap remains significant. In particular, for speech decoding, \citet{metzger2023neuroprosthesis} achieved a rate of 79 words per minute and reported a CER of 15.2\% on a dataset with 372 unique words, a vocabulary size comparable to ours.
\citet{willett2021handwriting} demonstrated typing speeds of 90 characters per minute with a CER below 6\% and an offline CER under 1\% when using a correction model. Both approaches rely on intracortical setups and require extensive recording sessions (11 hours per participant for \citet{willett2021handwriting}). Consequently an important research avenue will be to scale and adapt those tasks for MEG experiments. 

% Limits
Several challenges remain to be solved before the present method could be adapted to clinical applications. 
% real time
First, our model does not operate in real time. In particular, the transformer and language model here operate at the sentence level and thus require the trial to conclude before an output can be produced. 
In addition, the input of Brain2Qwerty requires the MEG segments to be aligned to keystrokes. Overall, a real-time architecture, akin to what is done with electromyography \citep{sivakumar2024emgqwerty} and speech recognition \citep{defossez2023decoding}, remains necessary to make the present proof-of-concept applicable in real time.  
% Second, due to the presence of subject-specific layers, our model cannot generalize to unseen subjects without some form of fine-tuning. 
% Eliminating the dependency on sentence-level correction, keystroke onsets and subject-specific layers are important endeavors towards real-time decoding. 
% patients

Second, our study was conducted exclusively with healthy participants and with a strictly supervised model. Training indeed requires knowing both the timing and identity of each character. While this setup may be suitable for patients with neurodegenerative conditions who may still possess  motor abilities, it is not applicable to locked-in individuals, who are completely unable to perform a typing task on a keyboard. Addressing this challenge may involve either adapting our typing task into an imagination task or designing AI systems capable of robust generalization across participants \citep{scotti2024mindeye2}.

Finally, while MEG outperforms EEG, current MEG systems, including the one used in the present study, are not wearable. This, however, may be resolved by the development of new MEG sensors based on optically pumped magnetometers (OPMs)~\citep{shah2013compact,schofield2022quantum,brickwedde2024applications}.

% Opening
% Finally, while our study demonstrates the feasibility of decoding typing, it does not offer a clear understanding of the underlying neural mechanisms driving these outcomes. This aspect is addressed in greater detail in a companion paper. 

Overall, the present results serve as a stepping stone toward developing safer and more accessible non-invasive brain-computer interfaces, ultimately enabling solutions for individuals who have partially or completely lost the ability to communicate.

\section{Methods} \label{section:decoder}
We aim to decode language production from non-invasive brain recordings. To achieve this, participants typed sentences on a keyboard while their brain activity was recorded using EEG or MEG. These two devices measure neural activity at a millisecond level, with EEG capturing electric fields and MEG detecting magnetic fields that are both generated by cortical neurons and recorded from sensors distributed across the scalp.

\subsection{Experimental Protocol}
\paragraph{Cohort. }
% We present a new dataset collected at the Basque Center on Cognition, Brain and Language (BCBL). Table~\ref{tab:dataset} summarizes the key characteristics of the dataset after preprocessing. 
We recruited 35 healthly adult volunteers to participate in our study at the Basque Center on Cognition, Brain and Language (BCBL) in Spain. This group was composed of 23\% of men and 77\% of women with an average age of 31.6$\pm$5.2 years. All participants were right-handed and skilled at typing. Participants had to type words they were hearing on a keyboard covered by a cardboard box. They were selected if their typing accuracy was above or equal to 80\%. They are all native Spanish speakers with no declared prior history of neurological or psychiatric disorders. Their brain activity was recorded with either EEG or MEG for 0.88$\pm$0.02 and 0.93$\pm$0.01 hours respectively, amounting to a total of 17.7 and 21.5 hours of typing. Five participants took part in both EEG and MEG sessions. One participant was excluded from the MEG study due to the presence of a metallic component during the recording. Participants gave their informed consent and were compensated 12 euros per hour for their participation. This study was approved by the local ethics committee. The same dataset is used to investigate the underlying neural mechanisms driving this task in a companion paper \citep{lucy2025}. 

\paragraph{Devices. }
% MEG system
The MEG system is a Megin system with 306 channels (102 magneto-meters and 204 planar gradiometers) recording at a sampling rate of 1\,kHz, with an online high-pass filter set at 0.1\,Hz and a low-pass filter at 330\,Hz.
% EEG system
The EEG system is an actiCAP slim from BrainVision\footnote{\url{https://brainvision.com/products/acticap-slim-acticap-snap/}}, with 64 channels (61 EEG channels and 3 ocular channels), a BRAINAMP DC amplifier, and sampled at 1\,kHz with an online high-pass filter set at 0.02\,Hz.

% The EEG system is composed of with 61 wet-electrodes, sampled at 1\,kHz with an online high-pass filter set at 0.02\,Hz. 
% Keyboard
Standard keyboards contain electronic and metallic parts that generate artifacts in the MEG. Consequently, we used a custom Magnetic-Resonance-compatible QWERTY keyboard from HybridMojo (LLC), and further modified it to replace the standard metallic springs with non-ferromagnetic silver-spring mechanisms.

\paragraph{Task. }
Participants were seated in front of a projected screen (100\,cm for MEG and 70\,cm for EEG away from their eyes), and with our custom keyboard placed on a stable platform. The distance between M/EEG sensors and the keyboard was 70\,cm.
% Participants were seated in front of a projected screen with our custom keyboard placed on a stable platform. 
This setup ensured participants could type in a natural position. Each trial consisted of three steps: read, wait, type. 
First, a sentence was presented on the screen, with a rapid serial visual presentation protocol (RSVP; i.e. one word at a time). Each word was presented in a black font, in all upper-case, on a 50\% gray background for a random duration between 465 and 665\,ms without intervals between words.
Second, after the disappearance of the last word of each sentence, a black fixation cross was displayed on the screen for 1.5 seconds. 
Third, the disappearance of the fixation cross signaled the start of the typing phase. No letters were presented on the screen during typing. Nevertheless, we added minimal visual feedback: a small black square at the center of the screen rotated clockwise by 10 degrees on every keystroke. This feedback ensured that eye movements were not correlated with linguistic features, as it is usually the case in left-to-right reading.
Each session consisted of two blocks of 64 sentences each. The first four sentences of each session were training sentences and were different from the 128 unique sentences in the protocol. During the first two training sentences, participants received visual feedback while typing. The other two sentences were used to train them on the task (typing with minimal visual feedback).

\paragraph{Instruction and stimuli. }
% instruction
Participants were instructed to type the sentence that was presented in RSVP as accurately as possible without using backspaces to correct errors and while fixating on the center of the screen. To avoid using diacritic marks that occur in Spanish (e.g., é, á, í, ó, ú, ü, and ñ), all words were presented in upper case, and participants were instructed to think of writing in upper case, and without accents. Participants pressed the return key at the end of the trial.
% sentences
Each session consisted of 128 unique Spanish sentences. All sentences were declarative Spanish sentences that contained between 5 and 8 words. They consisted of determiners, nouns, adjectives, prepositions and verbs. This led to a total of 4K sentences and 146K characters across participants for EEG, and 5.1K sentences and 193K characters for MEG. 
% The use of the backspace key was not allowed, preventing participants from correcting typographical errors. As a result, the typed sentences could contain mistakes. This setup enabled two definitions of a "correct" sentence: either the intended Spanish sentence that participants were instructed to type or the actual typed sentence, which could include errors.

\paragraph{MEG and EEG preprocessing. }
In our pipeline, the EEG and MEG recordings were bandpass filtered between 0.1 and 20\,Hz, and resampled to 50\,Hz, using MNE-Python's default parameters~\citep{gramfort2014mne}. These continuous recordings were then segmented into 0.5\,s time windows around each key press from -0.2\,s to +0.3\,s. We applied baseline correction by subtracting the average channel-wise value in the (-0.2, 0) interval from each window. To ensure that the data was on the same scale, a \texttt{RobustScaler} was applied across time from \texttt{scikit-learn} \citep{pedregosa2011scikit} followed by a clamping operation. The scaler removes the median and scales the data according to the interquantile range.

\paragraph{Text preprocessing. }
We removed sentences that contained strictly more than 10 typographical errors (less than 5\% for both EEG and MEG).
To determine whether the participants performed typographical errors, we first used the \texttt{SequenceMatcher} from \texttt{difflib }\footnote{https://docs.python.org/3/library/difflib.html} to align the typed sentence to the original sentence using a Gestalt pattern matching algorithm \citep{Ratcliff1988} that determines the minimal number of character-edits. This approach yields, for each keystroke event, two variables: the key pressed and the key that should have been pressed (target). Unless stated otherwise, all analyses are based on the target key. 
On average, participants typed 152.0$\pm$3.2 characters per minute, leading to an average sentence production time of 5.7$\pm$0.2\,s. 3.6 $\pm$ 0.7\% of the keystrokes were typographical errors.

\paragraph{Train/validation/test splits.}
To ensure that our AI model does not memorize sentences, we divided the unique sentences into train (80\%), validation (10\%) and test splits (10\%). As sentences may be similar (e.g. same sentence with an additional component), we adopted a maximally diverse splitting strategy by implementing a custom data splitter based on \texttt{sklearn}’s \texttt{AgglomerativeClustering} \citep{pedregosa2011scikit}. This method clusters unique sentences into groups of similar data points, where similarity is determined using Term Frequency-Inverse Document Frequency (TF-IDF) \citep{SparckJones1972} cosine similarity with a threshold of 0.5 (i.e., sentences with a similarity score above 0.5 are grouped in the same split). These clusters are then assigned to the training, validation and testing splits with relative ratios of 80/10/10. This approach prevents the model from artificially improving test performance by memorizing similar sentences across participants. 
Throughout the study, we presented our results using a fixed splitting seed, except in Fig. \ref{fig:performance_sentence} where we evaluated performance across sentences. For this figure, we aggregated results from multiple models with different splitting seeds to ensure broader sentence coverage, similar to a cross-validation approach.

\subsection{Decoder} \label{subsection:decoder}
The goal for our decoding model is to predict each keystroke based on 0.5\,s windows of M/EEG signals. Formally, the objective is to learn a mapping from brain recordings to class probabilities: $
f: \mathbb{R}^{s \times t} \rightarrow $[0,1]$^C
$, where \( s \) represents the number of M/EEG sensors, \( t \) denotes the number of M/EEG time samples in the window, and \( C \) is the number of available keys. We consider $C=29$ distinct classes, which include all the letters of the Latin alphabet as well as three special classes: one for space, another for numbers, and the last for all other special characters. 

% The model predicts \(\hat{Y}\), such that:

% \[
% \hat{Y} = f(X) \approx Y,
% \]

% where \( X \in \mathbb{R}^{c \times t} \) is the input and \( Y \in \mathbb{R}^C \) is the one-hot encoded ground truth. The cross-entropy loss is then used to minimize the difference between the predicted \(\hat{Y}\) and the true \(Y\), ensuring accurate classification.

\subsubsection{Architecture}
Our Brain2Qwerty model is composed of three successive modules (Fig. \ref{fig:approach}). 

\paragraph{Convolutional Module.}
The first building block is a modified convolutional model originally introduced by \cite{defossez2023decoding}. This model consists of four main parts. The first component employs a spatial attention mechanism to encode the relative positions of the sensors. The second introduces a subject-specific linear layer to account for differences between subjects. The third component is a convolutional neural network architecture, comprising 8 sequential blocks that employ a kernel size of 3 and a dilation period of 3, and incorporate skip connections, dropout regularization, and GELU activation functions. Finally, the temporal dimension is pooled with a single-head self-attention layer. Thus, for each window $\mathbf{X}\in \mathbb R^{s\times t}$, the CNN outputs $\mathbf{z}\in \mathbb R^{h}$, with $h=2,048$. For clarity, this module will hereafter be referred to as Convolutional Module (Conv).

\paragraph{Transformer Module.}
The outputs of the Convolutional Module are then input to a transformer Module (Trans). The receptive field of the transformer is restricted to a unique sentence: $\mathbf{Z}\in \mathbb R^{n\times h}$. The transformer is used to refine the keystroke predictions by exploiting contextual information and consists of 4 layers with 2 attention heads per layer, maintaining consistent input and output dimensions. Finally, a linear layer projects the transformer's outputs to obtain the logits of each character \(\mathbf{\tilde{Y}} \in  \mathbb{R}^{n\times C} \).

\paragraph{Language Model.}
Finally, the output of the transformer \(\mathbf{\tilde{Y}}\) is input to a language model, so as to leverage the statistical regularities of natural language. 
% pretraining
For this, we used a 9-gram character-level model constructed using the KenLM library \citep{KenLM} and pretrained on the Spanish Wikipedia Corpus \citep{wikidump}. This library optimizes both speed and memory efficiency by employing a prefix tree structure. 
%This design choice was inspired by the pipeline presented in \cite{sivakumar2024emgqwerty}, where EMG signals are used to decode typing. 
% Our goal is to develop a Spanish language-aware spelling mistake corrector that does not require extensive training or inference time, unlike character-level transformer models.
%
% inference
At inference time, the language model is input with a sequence of predicted characters, and causally predicts the most likely next character given the preceding predicted ones.

Formally, let \( i \in [1, 2, ..., n]\) denote the position of the character in a sequence of up to $n$ characters and $m=9$ be the order of the $m$-gram model. The probability of the next character \( \hat{c}_i \) given its preceding predictions \( \hat{c}_{i-m}, \ldots, \hat{c}_{i-1} \) is estimated as a weighted combination between the transformer's logits and the probabilities of the language model:

\[
\mathbf{P}(\hat{c}_i | \hat{c}_{i-m}, \ldots, \hat{c}_{i-1}) = \mathbf{P}_{\text{trans}}(\hat{c}_i) + \alpha \cdot \mathbf{P}_{\text{lm}}(\hat{c}_i | \hat{c}_{i-m}, \ldots, \hat{c}_{i-1}),
\]

where \( \mathbf P_{\text{trans}}(\hat{c}_i )= \log\left(\text{softmax} \left[(\mathbf{\tilde Y})\right]_i\right)\) represents the contribution from the core model, \( \mathbf P_{\text{lm}}(\hat{c}_i | \hat{c}_{i-m}, \ldots, \hat{c}_{i-1}) \) represents the probabilities from the language model, and \( \alpha \) is the language model weight. We use a beam search of size 30 and a language model weight of 5 found using a grid search, ensuring a tradeoff between inference time and decoding accuracy. This language modeling module outputs a sequence of $n$ characters that aim to regularize the predictions of the transformer with the statistics of natural language. Brain2Qwerty's final prediction is denoted as $\hat{\mathbf{Y}} \in \mathbb R^{n\times C}$. %Importantly, the language model does not alter the sequence length; it can only replace characters while preserving the original length.

\subsubsection{Training}
The Convolutional and Transformer modules are trained jointly with an unweighted cross-entropy loss in an end-to-end manner across all subjects, with the same hyperparameters for EEG and MEG recordings. This leads to a total of $\sim$ 400M parameters (258M for Conv, 138M for Trans). The model is trained for 100 epochs with a batch size of 128 using the AdamW optimizer \citep{loshchilov2019decoupledweightdecayregularization} with early stopping. We use the OneCycleLR scheduler \citep{smith2018superconvergencefasttrainingneural} (weight decay=$10^{-4}$; pct\_start=0.1), warming up the learning rate to $10^{-4}$ over the first 10 epochs then decaying linearly. Training was conducted on a single NVIDIA Tesla V100 Volta GPU with 32 GB of memory. The total runtime for training one model is $\sim$12 hours.

\subsubsection{Evaluation}

\paragraph{Hand Error Rate (HER).}
For analysis purposes and comparison with the classic BCI literature (e.g. \cite{lebedev2006brain}), we first consider HER. This metric estimates whether the target and the predicted characters correspond to the same left/right-hand split of the keyboard. Specifically, keys to the left of Y, H, and B are assigned to the left-hand category, while those to the right (including Y, H, and B) are assigned to the right-hand category. For this evaluation, special characters, numbers and space are excluded, as participants may use both hands to type them.

% \subsubsection{Character Accuracy}
% Character Accuracy evaluates the model's ability to correctly predict the typed character for each keystroke. For this metric, we use \textit{balanced accuracy} from the scikit-learn library \cite{pedregosa2011scikit}, which ensures fair performance evaluation across all 29 classes, regardless of class imbalance.

\paragraph{Character-error-rate (CER).}
CER is based on the Levenshtein distance which quantifies the minimum number of single-character edits required to transform the predicted sequence of keystrokes to the target sentence. 
A CER of 0 indicates perfect character-level accuracy. The formula for CER is given by \(\text{CER} = (s + d + a)\times\frac{1}{n}\), where \( s \), \( d \) and \( a \) stand for substitutions, deletions, and additions in the $n$-character sentence, respectively. %This metric provides a comprehensive measure of prediction errors, capturing all possible discrepancies between the predicted and reference text. 
Unless stated otherwise, we compute the CER at the sentence level using the $\texttt{Levenshtein}$\footnote{\url{https://pypi.org/project/Levenshtein/}} python library, and report the average CER across sentences.

% \subsubsection{Word Error Rate (WER)}
% WER is analogous to CER but operates at the word level. It measures the number of words in the predicted text that differ from those in the reference text, also utilizing the Levenshtein distance. \\

\subsubsection{Model comparison}

\paragraph{Statistics.}
For statistical comparisons, we employed the non-parametric tests provided by the \texttt{scipy} package~\citep{virtanen2020scipy}. For comparison across models within the same subjects, we used the Wilcoxon test. For comparison across subjects (e.g. for EEG versus MEG), we used the Mann Whitney U test. For time-course decoding, we further applied a false-discovery-rate (FDR) correction for multiple comparisons across time samples. 

\paragraph{Baseline Models.}
We consider two baseline models. 
% linear
The first one is a linear model implemented using the \texttt{RidgeClassifierCV} function from the scikit-learn library \citep{pedregosa2011scikit}, which was trained to predict characters from a single time sample of recording for each subject separately. The regularization parameter \( \alpha \) was selected through a nested cross-validation using a grid search logarithmic spanning from \( 10^{-2} \) to \( 10^8 \). This operation was repeated for each time sample between -0.5 and 0.5 seconds relative to the character onset. 

% EEGNet
As a second baseline, we need a model which uses the same setup as Brain2Qwerty, i.e. where all subjects are trained collectively and the temporal dimension is collapsed. We used EEGNet \citep{lawhern2018EEGNet}, a highly parameter-efficient model classically used in BCIs. We trained EEGNet with the same approach as our model. For our experiments, EEGNet is configured with a depth of 6 and a dropout rate of 0.3, which were selected based on a grid search over the validation set. EEGNet does not incorporate a subject-specific linear layer, which may compromise its performance in this particular setup where all subjects were trained collectively.
To compute the chance level while accounting for the imbalance across characters, we evaluated the performance of a dummy model that always predicts the most frequent character.

\paragraph{Companion paper.} We explore \emph{how} the brain produce a hierarchy of language representations in a companion paper \citep{lucy2025}. Note that Fig. \ref{fig:approach} (left) and Fig. \ref{fig:model_perf}B are shared between these two studies.

\section{Acknowledgments}
The authors would like to thank Maite Kaltzakorta, Manex Lete, Jessi Jacobsen, Daniel Nieto, Jone Iraeta, Araitz Garnika, Jaione Bengoetxea, Natalia Louleli, Naroa Miralles, Eñaut Zeberio, Craig Richter, Amets Esnal, and Olatz Andonegui, as well as Abhishek Charnalia and Pierre-Louis Xech for their critical help.
This research is supported by the Basque Government through the BERC 2022-2025 program and Funded by the Spanish State Research Agency through BCBL Severo Ochoa excellence accreditation CEX2020-001010/AEI/10.13039/501100011033.
Parts of this research were carried within the European Union's Horizon 2020 research and innovation programme under the Marie Skłodowska-Curie grant agreement No 945304 - Cofund AI4theSciences hosted by PSL University.

\clearpage
% \printbibliography  % biblatex not supported by template
\newpage
\bibliographystyle{assets/plainnat}
\bibliography{main}

\end{document}

%% file: figures/fig1.tex
\begin{figure*}[h]
    \centering
    \begin{subfigure}[b]{\linewidth}
        \includegraphics[width=\linewidth]{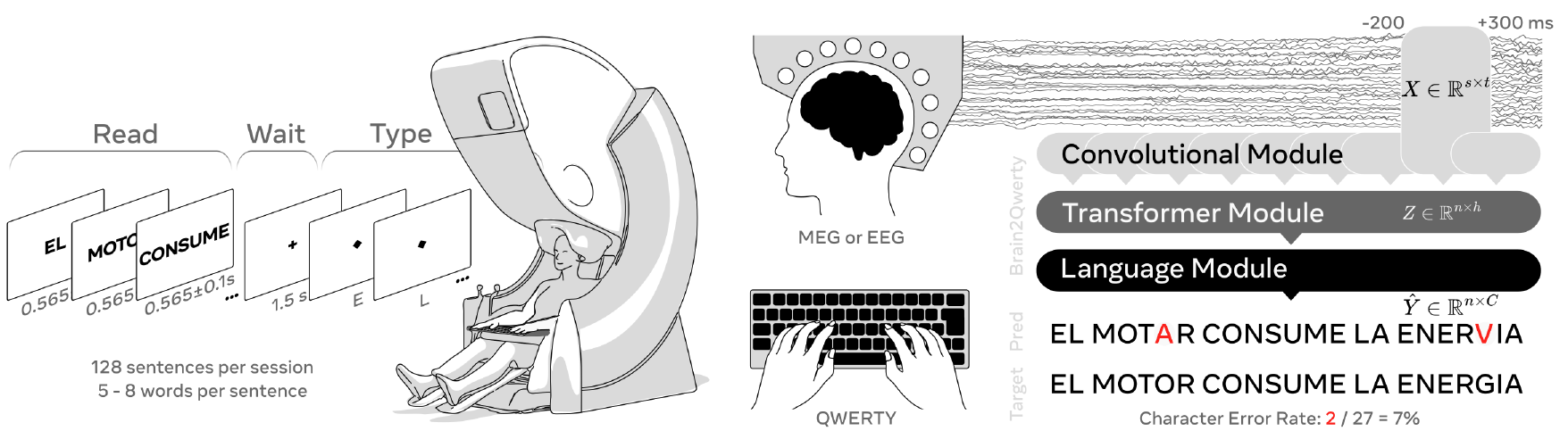}
    \end{subfigure}
    \caption{
        \textbf{Approach.}
        Recordings from 35 participants were obtained using electro-encephalography (EEG) and magneto-encephalography (MEG). Sentences were displayed word-by-word on a screen. Following the final word, a visual cue prompted them to begin typing this sentence, without visual feedback. Our Brain2Qwerty model includes three core stages to decode text from brain activity: (1) a convolutional module, input with 500\,ms windows of M/EEG signals, (2) a transformer module trained at the sentence level, and (3) a pretrained language model to correct the outputs of the transformer. Performance is assessed using a Character Error Rate (CER) at the sentence level. An analysis of \emph{how} the brain performs typing is described in a companion paper \citep{lucy2025}.
    }
    \label{fig:approach}
\end{figure*}

%% file: figures/fig2.tex
\begin{figure*}[!ht]
    \centering
    \vspace{-0.5cm}
    \includegraphics[width=1.\textwidth]{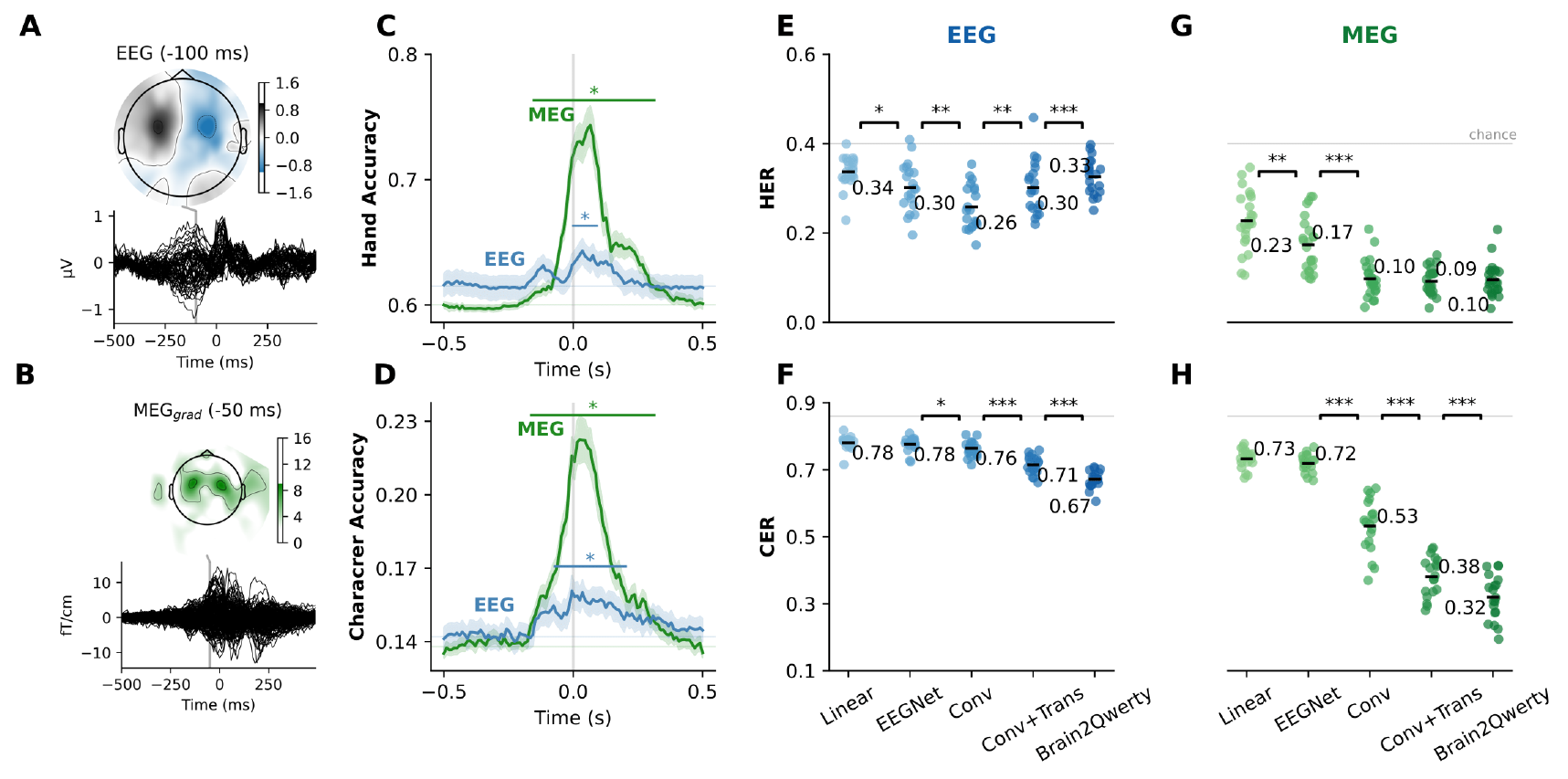}
    \caption{
        \textbf{Decoding Performance across models.}
        \textbf{A.} Difference in EEG evoked responses between left vs right hand key presses. Each black line is the differential voltage of a sensor relative to key press.  
        \textbf{B.} Same as A but for MEG.
        \textbf{C.} Linear classifiers are trained, at each time sample, to predict the left vs right hand relative to each key press. The gray line represents chance level and the error bar is the standard error of the mean across participants. Significant decoding scores (p < 0.05) are marked with a star. 
        \textbf{D.} Same as C but for character classification. 
        \textbf{E-H.} Comparison of baselines (linear and EEGNet), and ablation of our three-step Brain2Qwerty model (Conv+Trans+Language Model), for both hand-error-rate (HER) and character-error-rate (CER). Each point represents the average score of a single participant. \\
        Statistical significance is denoted with p < 0.05 (*), p < 0.01 (**), and p < 0.001 (***).
    }
    \label{fig:model_perf}
\end{figure*}

%% file: figures/fig3.tex
\begin{figure*}[htb] 
    \centering
    \includegraphics[width=1.\textwidth]{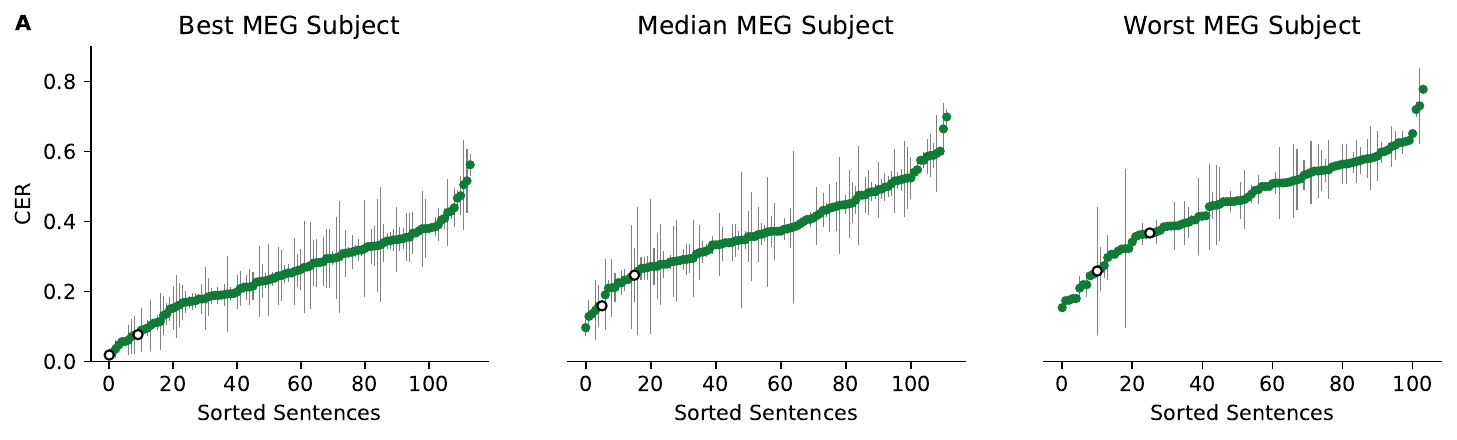} 
     \centering
     \begin{minipage}{0.49\textwidth}
        \centering
        
        \begin{tabular}{p{0.3\textwidth} p{0.7\textwidth}}
            \textbf{\small  B} \\ \\
            \textbf{True:} & \typing{las teorias reducen los numeros} \\
            \textbf{Best subject:} & \typing{\textcolor[rgb]{0.204, 0.596, 0.859}{las teorias reducen los numeros}} \\
            \textbf{Median subject:} & \typing{\textcolor[rgb]{0.204, 0.596, 0.859}{las teorias }\textcolor[rgb]{0.882, 0.071, 0.188}{exig}\textcolor[rgb]{0.204, 0.596, 0.859}{en los }\textcolor[rgb]{0.882, 0.071, 0.188}{homb}\textcolor[rgb]{0.204, 0.596, 0.859}{ros}} \\
            \textbf{Worst subject:} & \typing{\textcolor[rgb]{0.204, 0.596, 0.859}{las} \textcolor[rgb]{0.882, 0.071, 0.188}{ranc}\textcolor[rgb]{0.204, 0.596, 0.859}{ias re}\textcolor[rgb]{0.882, 0.071, 0.188}{vis}\textcolor[rgb]{0.204, 0.596, 0.859}{en los numer}\textcolor[rgb]{0.882, 0.071, 0.188}{ad}} \\
            \addlinespace

        \end{tabular}
    \end{minipage}
    \hfill
    \begin{minipage}{0.49\textwidth}
        \centering
        \begin{tabular}{p{0.3\textwidth} p{0.7\textwidth}}
            \textbf{\small  } \\ \\
            \textbf{True:} & \typing{la estadistica sigue la distribucion} \\
            \textbf{Best subject:} & \typing{\textcolor[rgb]{0.204, 0.596, 0.859}{la estadistica sigue la distribucion}} \\
            \textbf{Median subject:} & \typing{\textcolor[rgb]{0.882, 0.071, 0.188}{stamistosa} \textcolor[rgb]{0.204, 0.596, 0.859}{sigue la distribucion}} \\
            \textbf{Worst subject:} & \typing{\textcolor[rgb]{0.204, 0.596, 0.859}{la estadistica} \textcolor[rgb]{0.882, 0.071, 0.188}{f}\textcolor[rgb]{0.204, 0.596, 0.859}{igu}\textcolor[rgb]{0.882, 0.071, 0.188}{ra de petrilla lo}} \\

            \addlinespace
    % \textbf{True:} & \typing{el centro describe las parabolas }\\
    % \textbf{Type:} & \typing{\underline{w}l centro describe las parabolas }\\
    % \textbf{Decode:} & \typing{\textcolor[rgb]{0.882, 0.071, 0.188}{\underline{e}}\textcolor[rgb]{0.204, 0.596, 0.859}{l centro} \textcolor[rgb]{0.882, 0.071, 0.188}{trece de} \textcolor[rgb]{0.204, 0.596, 0.859}{las} \textcolor[rgb]{0.882, 0.071, 0.188}{carabin}\textcolor[rgb]{0.204, 0.596, 0.859}{as} }\\
    % \addlinespace
    % \textbf{True:} & \typing{los usos ofrecen las ventajas energeticas }\\
    % \textbf{Type:} & \typing{los usos ofrecen las ventajas energeticas }\\
    % \textbf{Decode:} & \typing{\textcolor[rgb]{0.204, 0.596, 0.859}{los} \textcolor[rgb]{0.882, 0.071, 0.188}{para me}\textcolor[rgb]{0.204, 0.596, 0.859}{recen las venta}\textcolor[rgb]{0.882, 0.071, 0.188}{n}\textcolor[rgb]{0.204, 0.596, 0.859}{as e}\textcolor[rgb]{0.882, 0.071, 0.188}{structur}\textcolor[rgb]{0.204, 0.596, 0.859}{as} }\\
    % \addlinespace
    % \textbf{True:} & \typing{la explicacion aclara la pregunta de la evaluacion }\\
    % \textbf{Type:} & \typing{la explicacion aclara la pregunta de la evaluacion }\\
    % \textbf{Decode:} & \typing{\textcolor[rgb]{0.204, 0.596, 0.859}{la explicacion}\textcolor[rgb]{0.882, 0.071, 0.188}{es para} \textcolor[rgb]{0.204, 0.596, 0.859}{la pre}\textcolor[rgb]{0.882, 0.071, 0.188}{sentan a} \textcolor[rgb]{0.204, 0.596, 0.859}{la }\textcolor[rgb]{0.882, 0.071, 0.188}{respirar}\textcolor[rgb]{0.204, 0.596, 0.859}{on} }\\
    \end{tabular}
    \end{minipage}

    \caption{
        \textbf{Sentence-level performance for Best, Median and Worst MEG subjects.} \\
        \textbf{A.} Character-error-rate for three representative subjects. Each dot represents a unique sentence, with error bars indicating the standard error of the mean across repetitions. White dots corresponds to the sentences displayed below. 
        \textbf{B.} Decoding predictions for two sentences.  %Brain2Qwerty achieves CERs of 0.06, 0.19, and 0.25 for the best, median and worst subjects in the first sentence (left) and 0, 0.23 and 0.4 in the second sentence (right). 
        Several splitting seeds were used to obtain the predictions across sentences.
    }
    \label{fig:performance_sentence} % Label for referencing the figure
\end{figure*}

%% file: figures/fig4.tex
\begin{figure*}[!ht] 
    \centering
    \includegraphics[width=1.\textwidth]{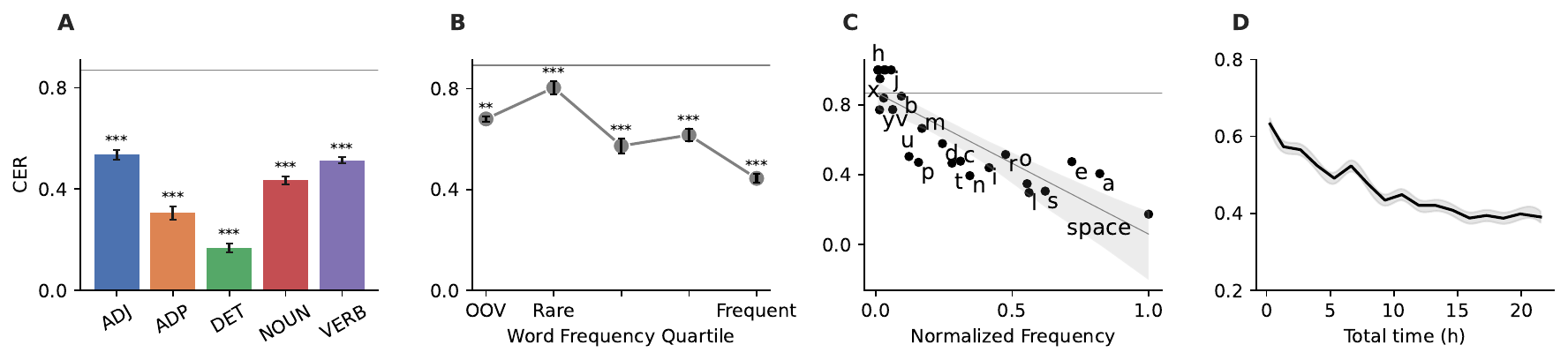} 
     \caption{
        \textbf{Analysis of character- and word-level performance.} \\
        The results presented are specific to MEG data processed using the Conv+Trans model.
        \textbf{A.} Character-error-rate (CER) is evaluated across different part-of-speech categories to evaluate how performance varies across adjectives (ADJ), nouns, verbs, determiners (DET), and prepositions (ADP). 
        \textbf{B.} CER as a function of word frequency. Out-of-vocabulary (OOV) decoding is used to test whether Brain2Qwerty can decode words absent from the training set. 
        \textbf{C.} CER as a function of character frequency. 
        \textbf{D.} CER as a function of recording time included in the training set. \\
    }
    \label{fig:model_analysis} % Label for referencing the figure
\end{figure*}

%% file: figures/fig5.tex
\begin{figure*}[!ht] 
    \centering
    \includegraphics[width=1.\textwidth]{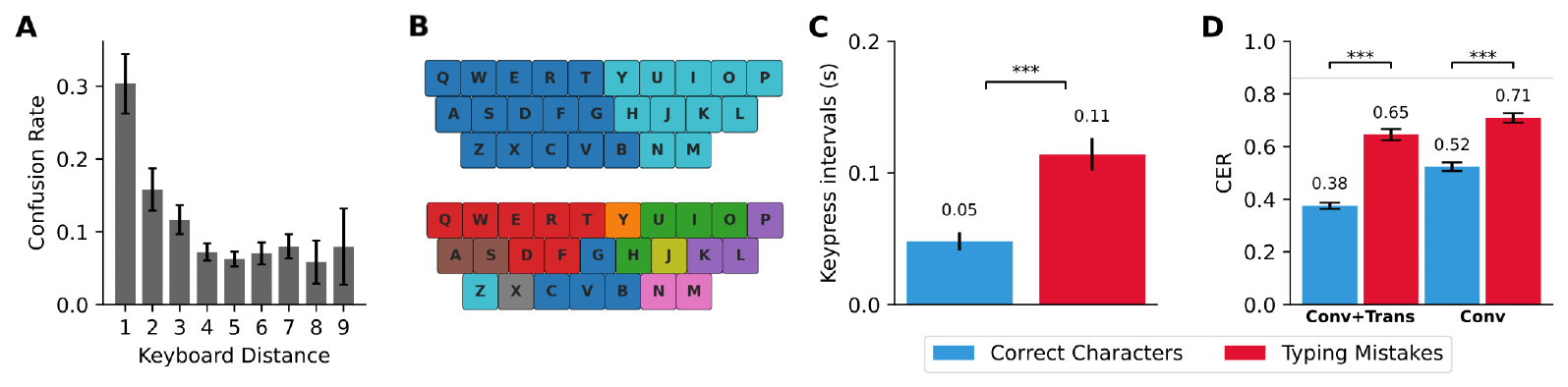} 
     \caption{
        \textbf{Impact of keyboard layout and typing errors.} \\
        The results presented are specific to MEG data processed using the Conv+Trans model.
        \textbf{A.} Keyboard Distance Effect. Confusion rate is analyzed against normalized keyboard distance. 
        \textbf{B.} Clustering Analysis. K-means clustering of the model embeddings with 2 (top) and 10 (bottom) clusters respectively. 
        \textbf{C.} Keypress Intervals Analysis. Comparison of keypress interval for correct keystrokes and typing errors, focusing on both preceding and subsequent characters. The sum of the two intervals is displayed. 
        \textbf{D.} Typing Mistakes Performance Differences. Performance comparison for correct characters versus typing errors using the Conv+Trans model (left) and the Conv model (right). 
    }
    \label{fig:motor_analysis} % Label for referencing the figure
\end{figure*}

%% file: main.bbl
\begin{thebibliography}{53}
\providecommand{\natexlab}[1]{#1}
\providecommand{\url}[1]{\texttt{#1}}
\expandafter\ifx\csname urlstyle\endcsname\relax
  \providecommand{\doi}[1]{doi: #1}\else
  \providecommand{\doi}{doi: \begingroup \urlstyle{rm}\Url}\fi

\bibitem[Abiri et~al.(2019)Abiri, Borhani, Sellers, Jiang, and Zhao]{abiri2019review}
R.~Abiri, S.~Borhani, E.~Sellers, Y.~Jiang, and X.~Zhao.
\newblock A comprehensive review of eeg-based brain-computer interface paradigms.
\newblock \emph{Journal of Neural Engineering}, 16\penalty0 (1):\penalty0 011001, Feb 2019.
\newblock \doi{10.1088/1741-2552/aaf12e}.
\newblock \url{https://doi.org/10.1088/1741-2552/aaf12e}.

\bibitem[Angrick et~al.(2019)Angrick, Herff, Mugler, Tate, Slutzky, Krusienski, and Schultz]{angrick2019speech}
M.~Angrick, C.~Herff, E.~Mugler, M.~C. Tate, M.~W. Slutzky, D.~J. Krusienski, and T.~Schultz.
\newblock Speech synthesis from ecog using densely connected 3d convolutional neural networks.
\newblock \emph{Journal of Neural Engineering}, 16\penalty0 (3):\penalty0 036019, Jun 2019.
\newblock \doi{10.1088/1741-2552/ab0c59}.
\newblock \url{https://doi.org/10.1088/1741-2552/ab0c59}.

\bibitem[Anumanchipalli et~al.(2019)Anumanchipalli, Chartier, and Chang]{anumanchipalli2019speech}
G.~K. Anumanchipalli, J.~Chartier, and E.~F. Chang.
\newblock Speech synthesis from neural decoding of spoken sentences.
\newblock \emph{Nature}, 568:\penalty0 493--498, 2019.
\newblock \doi{10.1038/s41586-019-1119-1}.
\newblock \url{https://doi.org/10.1038/s41586-019-1119-1}.

\bibitem[Baillet(2017)]{Baillet2017}
S.~Baillet.
\newblock Magnetoencephalography for brain electrophysiology and imaging.
\newblock \emph{Nature Neuroscience}, 20:\penalty0 327--339, 2017.
\newblock \doi{10.1038/nn.4504}.
\newblock \url{https://doi.org/10.1038/nn.4504}.

\bibitem[Baranauskas(2014)]{Baranauskas2014}
G.~Baranauskas.
\newblock What limits the performance of current invasive brain machine interfaces?
\newblock \emph{Frontiers in Systems Neuroscience}, 8:\penalty0 68, April 2014.

\bibitem[Bodien et~al.(2024)Bodien, Allanson, Cardone, et~al.]{bodien2024cognitive}
Y.~G. Bodien, J.~Allanson, P.~Cardone, et~al.
\newblock Cognitive motor dissociation in disorders of consciousness.
\newblock \emph{New England Journal of Medicine}, 391\penalty0 (7):\penalty0 598--608, Aug 2024.
\newblock \doi{10.1056/NEJMoa2400645}.
\newblock \url{https://doi.org/10.1056/NEJMoa2400645}.

\bibitem[Brickwedde et~al.(2024)Brickwedde, Anders, Kühn, Lofredi, Holtkamp, Kaindl, Grent-'t Jong, Krüger, Sander, and Uhlhaas]{brickwedde2024applications}
M.~Brickwedde, P.~Anders, A.~A. Kühn, R.~Lofredi, M.~Holtkamp, A.~M. Kaindl, T.~Grent-'t Jong, P.~Krüger, T.~Sander, and P.~J. Uhlhaas.
\newblock Applications of opm-meg for translational neuroscience: a perspective.
\newblock \emph{Translational Psychiatry}, 14\penalty0 (1):\penalty0 341, Aug 2024.
\newblock \doi{10.1038/s41398-024-03047-y}.
\newblock \url{https://doi.org/10.1038/s41398-024-03047-y}.

\bibitem[Bullard et~al.(2020)Bullard, Hutchison, Lee, Chestek, and Patil]{Bullard2020}
A.~Bullard, B.~Hutchison, J.~Lee, C.~Chestek, and P.~Patil.
\newblock Estimating risk for future intracranial, fully implanted, modular neuroprosthetic systems: A systematic review of hardware complications in clinical deep brain stimulation and experimental human intracortical arrays.
\newblock \emph{Neuromodulation: Technology at the Neural Interface}, 23\penalty0 (4):\penalty0 411--426, 2020.
\newblock ISSN 1094-7159.
\newblock \doi{10.1111/ner.13069}.
\newblock \url{https://doi.org/10.1111/ner.13069}.

\bibitem[Card et~al.(2024)Card, Wairagkar, Iacobacci, Hou, Singer-Clark, Willett, Kunz, Fan, Vahdati~Nia, Deo, Srinivasan, Choi, Glasser, Hochberg, Henderson, Shahlaie, Stavisky, and Brandman]{card2024accurate}
N.~S. Card, M.~Wairagkar, C.~Iacobacci, X.~Hou, T.~Singer-Clark, F.~R. Willett, E.~M. Kunz, C.~Fan, M.~Vahdati~Nia, D.~R. Deo, A.~Srinivasan, E.~Y. Choi, M.~F. Glasser, L.~R. Hochberg, J.~M. Henderson, K.~Shahlaie, S.~D. Stavisky, and D.~M. Brandman.
\newblock An accurate and rapidly calibrating speech neuroprosthesis.
\newblock \emph{New England Journal of Medicine}, 391\penalty0 (7):\penalty0 609--618, Aug 2024.
\newblock \doi{10.1056/NEJMoa2314132}.
\newblock \url{https://doi.org/10.1056/NEJMoa2314132}.

\bibitem[Cheng et~al.(2002)Cheng, Gao, Gao, and Xu]{cheng2002design}
M.~Cheng, X.~Gao, S.~Gao, and D.~Xu.
\newblock Design and implementation of a brain-computer interface with high transfer rates.
\newblock \emph{IEEE Transactions on Biomedical Engineering}, 49\penalty0 (10):\penalty0 1181--1186, Oct 2002.
\newblock \doi{10.1109/tbme.2002.803536}.
\newblock \url{https://doi.org/10.1109/tbme.2002.803536}.

\bibitem[Chevallier et~al.(2024)Chevallier, Carrara, Aristimunha, Guetschel, Sedlar, Lopes, Velut, Khazem, and Moreau]{chevallier2024}
S.~Chevallier, I.~Carrara, B.~Aristimunha, P.~Guetschel, S.~Sedlar, B.~Lopes, S.~Velut, S.~Khazem, and T.~Moreau.
\newblock The largest eeg-based bci reproducibility study for open science: the moabb benchmark, 2024.
\newblock \url{https://arxiv.org/abs/2404.15319}.

\bibitem[Chung et~al.(2019)Chung, Joo, Fan, Liu, Barnett, Chen, Geaghan-Breiner, Karlsson, Karlsson, Lee, Liang, Magland, Pebbles, Tooker, Greengard, Tolosa, and Frank]{chung2019highdensity}
J.~E. Chung, H.~R. Joo, J.~L. Fan, D.~F. Liu, A.~H. Barnett, S.~Chen, C.~Geaghan-Breiner, M.~P. Karlsson, M.~Karlsson, K.~Y. Lee, H.~Liang, J.~F. Magland, J.~A. Pebbles, A.~C. Tooker, L.~F. Greengard, V.~M. Tolosa, and L.~M. Frank.
\newblock High-density, long-lasting, and multi-region electrophysiological recordings using polymer electrode arrays.
\newblock \emph{Neuron}, 101\penalty0 (1):\penalty0 21--31.e5, Jan 2019.
\newblock \doi{10.1016/j.neuron.2018.11.002}.
\newblock \url{https://doi.org/10.1016/j.neuron.2018.11.002}.

\bibitem[Claassen et~al.(2019)Claassen, Doyle, Matory, Couch, Burger, Velazquez, Okonkwo, King, Park, Agarwal, Roh, Megjhani, Eliseyev, Connolly, and Rohaut]{Claassen2019}
J.~Claassen, K.~Doyle, A.~Matory, C.~Couch, K.~Burger, A.~Velazquez, J.~Okonkwo, J.-R. King, S.~Park, S.~Agarwal, D.~Roh, M.~Megjhani, A.~Eliseyev, E.~Connolly, and B.~Rohaut.
\newblock Detection of brain activation in unresponsive patients with acute brain injury.
\newblock \emph{New England Journal of Medicine}, 380\penalty0 (26):\penalty0 2497--2505, 2019.
\newblock \doi{10.1056/NEJMoa1812757}.
\newblock \url{https://doi.org/10.1056/NEJMoa1812757}.

\bibitem[Crell and Müller-Putz(2024)]{crell2024decoding}
M.~Crell and G.~Müller-Putz.
\newblock Handwritten character classification from eeg through continuous kinematic decoding.
\newblock \emph{Computers in Biology and Medicine}, 182:\penalty0 109132, 2024.
\newblock ISSN 0010-4825.
\newblock \doi{10.1016/j.compbiomed.2024.109132}.
\newblock \url{https://doi.org/10.1016/j.compbiomed.2024.109132}.

\bibitem[Donner et~al.(2009)Donner, Siegel, Fries, and Engel]{Donner2009}
T.~H. Donner, M.~Siegel, P.~Fries, and A.~K. Engel.
\newblock Buildup of choice-predictive activity in human motor cortex during perceptual decision making.
\newblock \emph{Current Biology}, 19\penalty0 (18):\penalty0 1581--1585, Sep 2009.
\newblock \doi{10.1016/j.cub.2009.07.066}.
\newblock \url{https://doi.org/10.1016/j.cub.2009.07.066}.

\bibitem[Défossez et~al.(2023)Défossez, Caucheteux, Rapin, Kabeli, and King]{defossez2023decoding}
A.~Défossez, C.~Caucheteux, J.~Rapin, O.~Kabeli, and J.-R. King.
\newblock Decoding speech perception from non-invasive brain recordings.
\newblock \emph{Nature Machine Intelligence}, 5:\penalty0 1097--1107, 2023.
\newblock \doi{10.1038/s42256-023-00714-5}.
\newblock \url{https://doi.org/10.1038/s42256-023-00714-5}.

\bibitem[Fekete et~al.(2023)Fekete, Zátonyi, Kaszás, et~al.]{Fekete2023}
Z.~Fekete, A.~Zátonyi, A.~Kaszás, et~al.
\newblock Transparent neural interfaces: challenges and solutions of microengineered multimodal implants designed to measure intact neuronal populations using high-resolution electrophysiology and microscopy simultaneously.
\newblock \emph{Microsystems \& Nanoengineering}, 9:\penalty0 66, 2023.
\newblock \doi{10.1038/s41378-023-00519-x}.
\newblock \url{https://doi.org/10.1038/s41378-023-00519-x}.

\bibitem[Goldenholz et~al.(2009)Goldenholz, Ahlfors, Hämäläinen, Sharon, Ishitobi, Vaina, and Stufflebeam]{Goldenholz2009}
D.~Goldenholz, S.~Ahlfors, M.~Hämäläinen, D.~Sharon, M.~Ishitobi, L.~Vaina, and S.~Stufflebeam.
\newblock Mapping the signal-to-noise ratios of cortical sources in magnetoencephalography and electroencephalography.
\newblock \emph{Human Brain Mapping}, 30\penalty0 (4):\penalty0 1077--1086, 2009.
\newblock \doi{10.1002/hbm.20571}.
\newblock \url{https://doi.org/10.1002/hbm.20571}.

\bibitem[Gramfort et~al.(2014)Gramfort, Luessi, Larson, Engemann, Strohmeier, Brodbeck, Parkkonen, and Hämäläinen]{gramfort2014mne}
A.~Gramfort, M.~Luessi, E.~Larson, D.~A. Engemann, D.~Strohmeier, C.~Brodbeck, L.~Parkkonen, and M.~S. Hämäläinen.
\newblock Mne software for processing meg and eeg data.
\newblock \emph{Neuroimage}, 86:\penalty0 446--460, Feb 2014.
\newblock \doi{10.1016/j.neuroimage.2013.10.027}.
\newblock \url{https://doi.org/10.1016/j.neuroimage.2013.10.027}.

\bibitem[Heafield(2011)]{KenLM}
K.~Heafield.
\newblock {K}en{LM}: Faster and smaller language model queries.
\newblock In Chris Callison-Burch, Philipp Koehn, Christof Monz, and Omar~F. Zaidan, editors, \emph{Proceedings of the Sixth Workshop on Statistical Machine Translation}, pages 187--197, Edinburgh, Scotland, July 2011. Association for Computational Linguistics.
\newblock \url{https://aclanthology.org/W11-2123/}.

\bibitem[Herff et~al.(2019)Herff, Diener, Angrick, Mugler, Tate, Goldrick, Krusienski, Slutzky, and Schultz]{herff2019generating}
C.~Herff, L.~Diener, M.~Angrick, E.~Mugler, M.~C. Tate, M.~A. Goldrick, D.~J. Krusienski, M.~W. Slutzky, and T.~Schultz.
\newblock Generating natural, intelligible speech from brain activity in motor, premotor, and inferior frontal cortices.
\newblock \emph{Frontiers in Neuroscience}, 13:\penalty0 1267, Nov 2019.
\newblock \doi{10.3389/fnins.2019.01267}.
\newblock \url{https://doi.org/10.3389/fnins.2019.01267}.

\bibitem[Hochberg et~al.(2012)Hochberg, Bacher, Jarosiewicz, et~al.]{Hochberg2012}
L.~Hochberg, D.~Bacher, B.~Jarosiewicz, et~al.
\newblock Reach and grasp by people with tetraplegia using a neurally controlled robotic arm.
\newblock \emph{Nature}, 485:\penalty0 372--375, 2012.
\newblock \doi{10.1038/nature11076}.
\newblock \url{https://doi.org/10.1038/nature11076}.

\bibitem[Hämäläinen et~al.(1993)Hämäläinen, Hari, Ilmoniemi, Knuutila, and Lounasmaa]{Hamalaainen1993}
M.~Hämäläinen, R.~Hari, R.~Ilmoniemi, J.~Knuutila, and O.~Lounasmaa.
\newblock Magnetoencephalography—theory, instrumentation, and applications to noninvasive studies of the working human brain.
\newblock \emph{Reviews of Modern Physics}, 65\penalty0 (2):\penalty0 413--497, 1993.
\newblock \doi{10.1103/RevModPhys.65.413}.
\newblock \url{https://doi.org/10.1103/RevModPhys.65.413}.

\bibitem[Jones(1972)]{SparckJones1972}
S.~Jones, K.
\newblock A statistical interpretation of term specificity and its application in retrieval.
\newblock \emph{Journal of Documentation}, 28\penalty0 (1):\penalty0 11--21, 1972.

\bibitem[Lawhern et~al.(2018)Lawhern, Solon, Waytowich, Gordon, Hung, and Lance]{lawhern2018EEGNet}
V.~J. Lawhern, A.~J. Solon, N.~R. Waytowich, S.~M. Gordon, C.~P. Hung, and B.~J. Lance.
\newblock Eegnet: a compact convolutional neural network for eeg-based brain-computer interfaces.
\newblock \emph{Journal of Neural Engineering}, 15\penalty0 (5):\penalty0 056013, Oct 2018.
\newblock \doi{10.1088/1741-2552/aace8c}.
\newblock \url{https://doi.org/10.1088/1741-2552/aace8c}.

\bibitem[Lebedev and Nicolelis(2006)]{lebedev2006brain}
M.~A. Lebedev and M.~A. Nicolelis.
\newblock Brain-machine interfaces: past, present and future.
\newblock \emph{Trends in Neurosciences}, 29\penalty0 (9):\penalty0 536--546, Sep 2006.
\newblock \doi{10.1016/j.tins.2006.07.004}.
\newblock \url{https://doi.org/10.1016/j.tins.2006.07.004}.

\bibitem[Leuthardt et~al.(2021)Leuthardt, Moran, and Mullen]{Leuthardt2021}
C.~Leuthardt, D.~Moran, W., and T.~Mullen.
\newblock Defining surgical terminology and risk for brain-computer interface technologies.
\newblock \emph{Frontiers in Neuroscience}, 15:\penalty0 599549, 2021.
\newblock \doi{10.3389/fnins.2021.599549}.
\newblock \url{https://doi.org/10.3389/fnins.2021.599549}.

\bibitem[Logan and Crump(2010)]{logan2010cognitive}
G.~D. Logan and M.~J.~C. Crump.
\newblock Cognitive illusions of authorship reveal hierarchical error detection in skilled typists.
\newblock \emph{Science}, 330\penalty0 (6004):\penalty0 683--686, 2010.
\newblock \doi{10.1126/science.1190483}.
\newblock \url{https://doi.org/10.1126/science.1190483}.

\bibitem[Loshchilov and Hutter(2019)]{loshchilov2019decoupledweightdecayregularization}
I.~Loshchilov and F.~Hutter.
\newblock Decoupled weight decay regularization, 2019.
\newblock \url{https://arxiv.org/abs/1711.05101}.

\bibitem[Mak and Wolpaw(2009)]{Mak2009}
J.~Mak and J.~Wolpaw.
\newblock Clinical applications of brain-computer interfaces: current state and future prospects.
\newblock \emph{IEEE Reviews in Biomedical Engineering}, 2:\penalty0 187--199, 2009.
\newblock \doi{10.1109/RBME.2009.2035356}.
\newblock \url{https://doi.org/10.1109/RBME.2009.2035356}.

\bibitem[Marchetti and Priftis(2014)]{marchetti2014effectiveness}
M.~Marchetti and K.~Priftis.
\newblock Effectiveness of the p3-speller in brain-computer interfaces for amyotrophic lateral sclerosis patients: a systematic review and meta-analysis.
\newblock \emph{Frontiers in Neuroengineering}, 7:\penalty0 12, May 2014.
\newblock \doi{10.3389/fneng.2014.00012}.
\newblock \url{https://doi.org/10.3389/fneng.2014.00012}.

\bibitem[Metzger et~al.(2022)Metzger, Liu, Moses, Dougherty, Seaton, Littlejohn, Chartier, Anumanchipalli, Tu-Chan, Ganguly, and Chang]{Metzger2022}
S.~L. Metzger, J.~R. Liu, D.~A. Moses, M.~E. Dougherty, M.~P. Seaton, K.~T. Littlejohn, J.~Chartier, G.~K. Anumanchipalli, A.~Tu-Chan, K.~Ganguly, and E.~F. Chang.
\newblock Generalizable spelling using a speech neuroprosthesis in an individual with severe limb and vocal paralysis.
\newblock \emph{Nature Communications}, 13\penalty0 (1):\penalty0 6510, Nov 2022.
\newblock \doi{10.1038/s41467-022-33611-3}.
\newblock \url{https://doi.org/10.1038/s41467-022-33611-3}.

\bibitem[Metzger et~al.(2023)Metzger, Littlejohn, Silva, Moses, Seaton, Wang, Dougherty, Liu, Wu, Berger, Zhuravleva, Tu-Chan, Ganguly, Anumanchipalli, and Chang]{metzger2023neuroprosthesis}
S.~L. Metzger, K.~T. Littlejohn, A.~B. Silva, D.~A. Moses, M.~P. Seaton, R.~Wang, M.~E. Dougherty, J.~R. Liu, P.~Wu, M.~A. Berger, I.~Zhuravleva, A.~Tu-Chan, K.~Ganguly, G.~K. Anumanchipalli, and E.~F. Chang.
\newblock A high-performance neuroprosthesis for speech decoding and avatar control.
\newblock \emph{Nature}, 620\penalty0 (7976):\penalty0 1037--1046, Aug 2023.
\newblock \doi{10.1038/s41586-023-06443-4}.
\newblock \url{https://doi.org/10.1038/s41586-023-06443-4}.

\bibitem[Moses et~al.(2021)Moses, Metzger, Liu, Anumanchipalli, Makin, Sun, Chartier, Dougherty, Liu, Abrams, Tu-Chan, Ganguly, and Chang]{moses2021neuroprosthesis}
D.~A. Moses, S.~L. Metzger, J.~R. Liu, G.~K. Anumanchipalli, J.~G. Makin, P.~F. Sun, J.~Chartier, M.~E. Dougherty, P.~M. Liu, G.~M. Abrams, A.~Tu-Chan, K.~Ganguly, and E.~F. Chang.
\newblock Neuroprosthesis for decoding speech in a paralyzed person with anarthria.
\newblock \emph{New England Journal of Medicine}, 385\penalty0 (3):\penalty0 217--227, Jul 2021.
\newblock \doi{10.1056/NEJMoa2027540}.
\newblock \url{https://doi.org/10.1056/NEJMoa2027540}.

\bibitem[Owen et~al.(2006)Owen, Coleman, Boly, Davis, Laureys, and Pickard]{owen2006detecting}
A.~M. Owen, M.~R. Coleman, M.~Boly, M.~H. Davis, S.~Laureys, and J.~D. Pickard.
\newblock Detecting awareness in the vegetative state.
\newblock \emph{Science}, 313\penalty0 (5792):\penalty0 1402, Sep 2006.
\newblock \doi{10.1126/science.1130197}.
\newblock \url{https://doi.org/10.1126/science.1130197}.

\bibitem[Pedregosa et~al.(2018)Pedregosa, Varoquaux, Gramfort, Michel, Thirion, Grisel, Blondel, Müller, Nothman, Louppe, Prettenhofer, Weiss, Dubourg, Vanderplas, Passos, Cournapeau, Brucher, Perrot, and Duchesnay]{pedregosa2011scikit}
F.~Pedregosa, G.~Varoquaux, A.~Gramfort, V.~Michel, B.~Thirion, O.~Grisel, M.~Blondel, A.~Müller, J.~Nothman, G.~Louppe, P.~Prettenhofer, R.~Weiss, V.~Dubourg, J.~Vanderplas, A.~Passos, D.~Cournapeau, M.~Brucher, M.~Perrot, and É. Duchesnay.
\newblock Scikit-learn: Machine learning in python, 2018.
\newblock \url{https://arxiv.org/abs/1201.0490}.

\bibitem[Pinet and Nozari(2020)]{Pinet2020}
S.~Pinet and N.~Nozari.
\newblock Electrophysiological correlates of monitoring in typing with and without visual feedback.
\newblock \emph{Journal of Cognitive Neuroscience}, 32\penalty0 (4):\penalty0 603--620, April 2020.
\newblock \doi{10.1162/jocn_a_01500}.
\newblock \url{https://doi.org/10.1162/jocn_a_01500}.

\bibitem[Ratcliff and Metzener(1988)]{Ratcliff1988}
W.~Ratcliff, J. and E.~Metzener, D.
\newblock Pattern matching: The gestalt approach.
\newblock \emph{Dr. Dobb's Journal}, page~46, July 1988.
\newblock \url{https://www.drdobbs.com/database/pattern-matching-the-gestalt-approach/184407970?pgno=5}.

\bibitem[Schofield et~al.(2022)Schofield, Boto, Shah, Hill, Osborne, Rea, Doyle, Holmes, Bowtell, Woolger, and Brookes]{schofield2022quantum}
H.~Schofield, E.~Boto, V.~Shah, R.~M. Hill, J.~Osborne, M.~Rea, C.~Doyle, N.~Holmes, R.~Bowtell, D.~Woolger, and M.~J. Brookes.
\newblock Quantum enabled functional neuroimaging: the why and how of magnetoencephalography using optically pumped magnetometers.
\newblock \emph{Contemporary Physics}, 63\penalty0 (3):\penalty0 161--179, 2022.
\newblock \doi{10.1080/00107514.2023.2182950}.
\newblock \url{https://doi.org/10.1080/00107514.2023.2182950}.

\bibitem[Scotti et~al.(2024)Scotti, Tripathy, Villanueva, Kneeland, Chen, Narang, Santhirasegaran, Xu, Naselaris, Norman, and Abraham]{scotti2024mindeye2}
S.~Scotti, P., M.~Tripathy, K.~T. Villanueva, C., R.~Kneeland, T.~Chen, A.~Narang, C.~Santhirasegaran, J.~Xu, T.~Naselaris, A.~Norman, K., and M.~Abraham, T.
\newblock Mindeye2: Shared-subject models enable fmri-to-image with 1 hour of data, 2024.
\newblock \url{https://arxiv.org/abs/2403.11207}.

\bibitem[Shah and Wakai(2013)]{shah2013compact}
V.~K. Shah and R.~T. Wakai.
\newblock A compact, high performance atomic magnetometer for biomedical applications.
\newblock \emph{Physics in Medicine and Biology}, 58\penalty0 (22):\penalty0 8153--8161, Nov 2013.
\newblock \doi{10.1088/0031-9155/58/22/8153}.
\newblock \url{https://doi.org/10.1088/0031-9155/58/22/8153}.

\bibitem[Sivakumar et~al.(2024)Sivakumar, Seely, Du, Bittner, Berenzweig, Bolarinwa, Gramfort, and Mandel]{sivakumar2024emgqwerty}
V.~Sivakumar, J.~Seely, A.~Du, R.~Bittner, S., A.~Berenzweig, A.~Bolarinwa, A.~Gramfort, and I.~Mandel, M.
\newblock emg2qwerty: A large dataset with baselines for touch typing using surface electromyography, 2024.
\newblock \url{https://arxiv.org/abs/2410.20081}.

\bibitem[Smith and Topin(2018)]{smith2018superconvergencefasttrainingneural}
N.~Smith, L. and N.~Topin.
\newblock Super-convergence: Very fast training of neural networks using large learning rates, 2018.
\newblock \url{https://arxiv.org/abs/1708.07120}.

\bibitem[Tang et~al.(2023)Tang, LeBel, Jain, and Huth]{tang2023semantic}
J.~Tang, A.~LeBel, S.~Jain, and A.~G. Huth.
\newblock Semantic reconstruction of continuous language from non-invasive brain recordings.
\newblock \emph{Nature Neuroscience}, 26\penalty0 (5):\penalty0 858--866, May 2023.
\newblock \doi{10.1038/s41593-023-01304-9}.
\newblock \url{https://doi.org/10.1038/s41593-023-01304-9}.

\bibitem[Virtanen et~al.(2020)Virtanen, Gommers, Oliphant, et~al.]{virtanen2020scipy}
P.~Virtanen, R.~Gommers, T.~Oliphant, et~al.
\newblock Scipy 1.0: fundamental algorithms for scientific computing in python.
\newblock \emph{Nature Methods}, 17\penalty0 (3):\penalty0 261–272, Feb 2020.
\newblock \doi{10.1038/s41592-019-0686-2}.
\newblock \url{http://dx.doi.org/10.1038/s41592-019-0686-2}.

\bibitem[Wairagkar et~al.(2024)Wairagkar, Card, Singer-Clark, Hou, Iacobacci, Hochberg, Brandman, and Stavisky]{wairagkar2024voice}
M.~Wairagkar, N.~S. Card, T.~Singer-Clark, X.~Hou, C.~Iacobacci, L.~R. Hochberg, D.~M. Brandman, and S.~D. Stavisky.
\newblock An instantaneous voice synthesis neuroprosthesis, Sep 2024.
\newblock \url{https://doi.org/10.1101/2024.08.14.607690}.
\newblock bioRxiv [Preprint].

\bibitem[Wikimedia(2005)]{wikidump}
Wikimedia.
\newblock Wikimedia downloads, 2005.
\newblock \url{https://dumps.wikimedia.org}.

\bibitem[Willett et~al.(2021)Willett, Avansino, Hochberg, Henderson, and Shenoy]{willett2021handwriting}
F.~R. Willett, D.~T. Avansino, L.~R. Hochberg, J.~M. Henderson, and K.~V. Shenoy.
\newblock High-performance brain-to-text communication via handwriting.
\newblock \emph{Nature}, 593\penalty0 (7858):\penalty0 249--254, May 2021.
\newblock \doi{10.1038/s41586-021-03506-2}.
\newblock \url{https://doi.org/10.1038/s41586-021-03506-2}.

\bibitem[Willett et~al.(2023)Willett, Kunz, Fan, Avansino, Wilson, Choi, Kamdar, Glasser, Hochberg, Druckmann, Shenoy, and Henderson]{willett2023neuroprosthesis}
F.~R. Willett, E.~M. Kunz, C.~Fan, D.~T. Avansino, G.~H. Wilson, E.~Y. Choi, F.~Kamdar, M.~F. Glasser, L.~R. Hochberg, S.~Druckmann, K.~V. Shenoy, and J.~M. Henderson.
\newblock A high-performance speech neuroprosthesis.
\newblock \emph{Nature}, 620\penalty0 (7976):\penalty0 1031--1036, Aug 2023.
\newblock \doi{10.1038/s41586-023-06377-x}.
\newblock \url{https://doi.org/10.1038/s41586-023-06377-x}.

\bibitem[Yasar et~al.(2024)Yasar, Gombkoto, Vyssotski, et~al.]{Yasar2024}
T.~Yasar, P.~Gombkoto, A.~Vyssotski, et~al.
\newblock Months-long tracking of neuronal ensembles spanning multiple brain areas with ultra-flexible tentacle electrodes.
\newblock \emph{Nature Communications}, 15:\penalty0 4822, 2024.
\newblock \doi{10.1038/s41467-024-49226-9}.
\newblock \url{https://doi.org/10.1038/s41467-024-49226-9}.

\bibitem[Yi et~al.(2014)Yi, Qiu, Wang, Qi, Zhang, Zhou, He, and Ming]{Yi2014}
W.~Yi, S.~Qiu, K.~Wang, H.~Qi, L.~Zhang, P.~Zhou, F.~He, and D.~Ming.
\newblock Evaluation of eeg oscillatory patterns and cognitive process during simple and compound limb motor imagery.
\newblock \emph{PLoS One}, 9\penalty0 (12):\penalty0 e114853, Dec 2014.
\newblock \doi{10.1371/journal.pone.0114853}.
\newblock \url{https://doi.org/10.1371/journal.pone.0114853}.

\bibitem[Zhang et~al.(2025)Zhang, Lévy, d'Ascoli, Rapin, Alario, Bourdillon, Pinet, and King]{lucy2025}
M.~Zhang, J.~Lévy, S.~d'Ascoli, J.~Rapin, F.-X. Alario, P.~Bourdillon, S.~Pinet, and J.-R. King.
\newblock From thought to action: How the hierarchy of neural dynamics supports language production.
\newblock 2025.

\bibitem[Zhou et~al.(2024)Zhou, Tian, Li, et~al.]{Zhou2024}
C.~Zhou, Y.~Tian, G.~Li, et~al.
\newblock Through-polymer, via technology-enabled, flexible, lightweight, and integrated devices for implantable neural probes.
\newblock \emph{Microsystems \& Nanoengineering}, 10:\penalty0 54, 2024.
\newblock \doi{10.1038/s41378-024-00691-8}.
\newblock \url{https://doi.org/10.1038/s41378-024-00691-8}.

\end{thebibliography}
